\documentclass[]{emulateapj}
\usepackage{epsfig, natbib, graphicx, color}
\usepackage{apjfonts}

\input epsf

\newcommand{\Mpc}{\mbox{Mpc}}
\newcommand{\hMpc}{h^{-1}\mbox{Mpc}}
\newcommand{\msun}{M_\odot}
\newcommand{\bm}[1]{\mathbf{#1}}

\newcommand{\avg}[1]{\left\langle #1 \right\rangle}

\newcommand{\bx}{\bm{x}}
\newcommand{\bp}{\bm{p}}
\newcommand{\bq}{\bm{q}}
\newcommand{\br}{\bm{r}}

\newcommand{\kpc}{\mbox{kpc}}

\newcommand{\Lx}{L_X}

\newcommand{\keV}{\mbox{keV}}

\newcommand{\erf}{\mbox{erf}}

\shortauthors{Rozo et al.}
\shorttitle{Scatter in the Mass-Richness Relation}

\begin{document}
\title{Constraining the Scatter in the Mass-Richness Relation of maxBCG Clusters With Weak Lensing
and X-ray Data}
\author{Eduardo Rozo\altaffilmark{1}, Eli S. Rykoff\altaffilmark{2}, August Evrard\altaffilmark{3,4,5}, Matthew Becker\altaffilmark{6},
Timothy McKay\altaffilmark{3,4,5}, Risa H. Wechsler\altaffilmark{7}, Benjamin P. Koester\altaffilmark{8,9},
Jiangang Hao\altaffilmark{3}, Sarah Hansen\altaffilmark{8,9}, 
Erin Sheldon\altaffilmark{10}, David Johnston\altaffilmark{11}, James Annis\altaffilmark{12}, Joshua Frieman\altaffilmark{8,9,12}}

\altaffiltext{1}{Center for Cosmology and Astro-Particle Physics (CCAPP), The Ohio State University, Columbus, OH 43210}
\altaffiltext{2}{TABASGO Fellow, Physics Department, University of California at Santa Barbara, 2233B Broida Hall, Santa Barbara, CA 93106}
\altaffiltext{3}{Physics Department, University of Michigan, Ann Arbor, MI 48109}
\altaffiltext{4}{Astronomy Department, Universityof Michigan, AnnArbor, MI 48109}
\altaffiltext{5}{Michigan Center for Theoretical Physics, Ann Arbor, MI 48109}
\altaffiltext{6}{Department of Physics, The University of Chicago, Chicago, IL 60637}
\altaffiltext{7}{Kavli Institute for Particle Astrophysics \& Cosmology,
  Physics Department, and Stanford Linear Accelerator Center,
  Stanford University, Stanford, CA 94305}
\altaffiltext{8}{Department of Astronomy and Astrophysics, The University of Chicago, Chicago, IL 60637}
\altaffiltext{9}{Kavli Institute for Cosmological Physics, The University of Chicago, Chicago, IL 60637} 
\altaffiltext{10}{Center for Cosmology and Particle Physics, Physics Department, New York University, New York, NY 10003}
\altaffiltext{11}{ Jet Propulsion Laboratory, 4800 Oak Grove Drive, Pasadena, CA 91109}
\altaffiltext{12}{Fermi National Accelerator Laboratory, P.O. Box500, Batavia, IL 60510}

\begin{abstract}
We measure the logarithmic scatter in mass at fixed richness for clusters in the maxBCG cluster catalog, an
optically selected cluster sample drawn from SDSS imaging data.  Our measurement is achieved by demanding
consistency between available weak lensing
and X-ray measurements of the maxBCG clusters, and the X-ray luminosity--mass relation inferred from the
400d X-ray cluster survey, a flux limited X-ray cluster survey.
We find $\sigma_{\ln M|N_{200}}=0.45^{+0.20}_{-0.18}$
($95\%$ CL) at $N_{200}\approx 40$, where $N_{200}$ is the number of red sequence galaxies in
a cluster.  As a byproduct of our analysis, we also obtain a constraint
on the correlation coefficient between $\ln \Lx$ and $\ln M$ at fixed richness, which is best expressed
as a lower limit, $r_{L,M|N} \geq 0.85\ (95\%\ \mbox{CL})$.  This is the first observational constraint
placed on a correlation coefficient involving two different cluster mass tracers.
We use our results to produce a state of the art estimate of the halo mass function at $z=0.23$ --- the
median redshift of the maxBCG cluster sample --- 
and find that it is consistent with the WMAP5 cosmology.   Both the mass function data
and its covariance matrix are presented.
\end{abstract}

 \keywords{galaxies: clusters -- X-rays: galaxies: clusters - cosmology: observation}

\section{Introduction}

The space density of galaxy clusters as a function of cluster mass is a well-known
cosmological probe \citep[see e.g.][]{holderetal01,haimanetal01,rozoetal04,limahu04}, 
and ranks among the best observational tools
for constraining $\sigma_8$, 
the normalization of the matter power spectrum in the low redshift
universe \citep[see e.g.][]{frenketal90,henry91,
schueckeretal03,gladdersetal07,rozoetal07a}.\footnote{$\sigma_8$ 
is formally defined
as the variance of the linear matter density averaged over spheres with radius
$R=8\ h^{-1}\ \Mpc$.}  The basic idea is this: in the high mass limit,
the cluster mass function falls off exponentially with mass, with the fall-off  
depending sensitively on the amplitude of the matter density fluctuations.  
Observing this exponential cutoff can thus place tight constraints 
on $\sigma_8$.
In practice, however, the same exponential dependence that makes cluster abundances a powerful
cosmological probe also renders it susceptible to an important systematic effect,
namely uncertainties in the estimated masses of clusters.  

Because mass is not
a direct observable, cluster masses must be determined using observable mass
tracers such as X-ray emission, SZ decrements, weak lensing shear, or cluster richness (a measure of the galaxy
content of the cluster).   Of course, such mass estimators are noisy,
meaning there can be significant scatter between the observable mass tracer
and cluster mass.  Since the mass function declines steeply with mass, up-scattering of low mass 
systems into high mass bins can result in a significant boost to the number of systems with apparently 
high mass \citep{limahu05}.  If this effect is not properly modeled,
the value of $\sigma_8$ derived from such a cluster sample will be overestimated.

One approach for dealing with this difficulty is to employ mass tracers
that have minimal scatter, thereby reducing the impact of
said scatter on the recovered halo mass function. 
For instance, \citet{kravtsovetal06} introduced a new X-ray mass estimator, $Y_X=M_{gas}T_X$,
which in their simulations exhibits an intrinsic scatter of only $\approx 8\%$, independent
of the dynamical state of the cluster. 
Use of a mass estimator with such low scatter should lead to 
improved estimates of $\sigma_8$ from X-ray cluster surveys  \citep{pierpaolietal01,rb02,schueckeretal03,
henry04,staneketal06}.

Such tightly-correlated mass tracers are not always available.
In such cases, determination of the scatter in the mass-observable relation 
is critical to accurately inferring the mass function and thereby determining cosmological parameters.
Of course, in practice, it is impossible to determine this scatter to arbitrary
accuracy, but since the systematic boost to the mass function is
proportional to the square of the scatter \citep{limahu05} (i.e. the variance), even moderate constraints
on the scatter can result in tight $\sigma_8$ constraints.

In this paper, we use optical and X-ray observations to constrain the scatter in the mass--richness 
relation for the
maxBCG cluster catalog presented in \citet{koesteretal07a}.
Specifically, we use observational constraints on the mean 
mass--richness relation, and on the mean and scatter of the $\Lx-$richness relation,
to convert independent estimates of the scatter in the $\Lx-M$ relation into estimates
of the scatter in the mass--richness relation.  An interesting byproduct
of our analysis is a constraint on the correlation coefficient between 
mass and X-ray luminosity at fixed richness.  To our knowledge, this is the first
time that a correlation coefficient involving multiple cluster mass tracers
has been empirically determined.

The layout of the paper is as follows. In section \ref{sec:notation} we lay out the notation and definitions 
used throughout the paper.  Section \ref{sec:data} presents the
data sets used in our analysis.  In section \ref{sec:rough} we present a pedagogical
description of our method for constraining the scatter in the richness-mass relation,
while section \ref{sec:formalism} formalizes the argument.  Our results are found
in section \ref{sec:results}, and we compare them to previous work in section \ref{sec:other_work}.
In section \ref{sec:mf}, we use our result to estimate the halo mass function in the local
universe at $z=0.23$, the median redshift of the maxBCG cluster sample, and we demonstrate
that our recovered mass function is consistent with the latest cosmological constraints
from WMAP \citep{wmap08}.  A detailed cosmological analysis of our results will be presented
in a forthcoming paper (Rozo et al., in preparation).  Our summary and conclusions are 
presented in section \ref{sec:conclusions}.

\subsection{Notation and Conventions}
\label{sec:notation}

We summarize here the notation and conventions
employed in this work.
Given any three cluster mass tracers (possibly including mass itself) $X,Y,$
and $Z$, we make the standard assumption that the probability distribution $P(X,Y|Z)$ is 
a bivariate lognormal.  The parameters $A_{X|Z}$, $B_{X|Z}$, and $\alpha_{X|Z}$ are defined such that
\begin{eqnarray}
\avg{\ln X|Z} & = & A_{X|Z}+\alpha_{X|Z}\ln Z \\
\ln \avg{X|Z} & = & B_{X|Z} + \alpha_{X|Z}\ln Z.
\end{eqnarray}
Note the slopes of the mean and logarithmic mean are the same, as appropriate
for a log-normal distribution.  The scatter in $\ln X$ at fixed $Z$ is denoted $\sigma_{X|Z}$,
and the correlation coefficient between $\ln X$ and $\ln Y$ at fixed $Z$ is denoted
$r_{X,Y|Z}$.  {\it We emphasize that all quoted scatters are the scatter in the natural logarithm,
not in dex.}
Note these parameters are simply the elements of the covariance matrix
specifying the Gaussian distribution $P(\ln X,\ln Y|\ln Z)$.
Under our lognormal assumption for $P(X,Y|Z)$, the parameters $A_{X|Z}$ and 
$B_{X|Z}$ are related via
\begin{equation}
B_{X|Z} = A_{X|Z}+\frac{1}{2}\sigma_{X|Z}^2.
\end{equation}

In this work, the quantities of interest are cluster mass $M$, X-ray luminosity $\Lx$, and 
cluster richness $N$.  Unless otherwise specified, cluster mass is defined as 
$M_{500c}$, the mass contained within an overdensity of 
500 relative to critical.  $\Lx$ is the total luminosity in the rest-frame $0.5-2.0\ \keV$ band,
and $N$ is the maxBCG richness measure $N_{200}$, the number of red sequence galaxies
with luminosity above $0.4L_*$ within an aperture such that the mean density within said
radius is, on average, $200\Omega_m^{-1}$ times the mean galaxy density assuming $\Omega_m=0.3$.  
Likewise, unless otherwise
stated all parameters governing the relations between $M$, $\Lx$, and $N$ assume that
$M$ is measured in units of $10^{14}\ \msun$, $\Lx$ is measured in units of
$10^{43}\ \mathrm{ergs}/\mathrm{s}$, and $N$ is measured in ``units'' of $40$ galaxies.
For instance, including units explicitly, the mean relation between
cluster mass and richness reads
\begin{equation}
\frac{\avg{M|N}}{10^{14}\ \msun} = \exp(B_{M|N}) \left (\frac{N}{40}\right)^{\alpha_{M|N}}.
\end{equation}
A Hubble constant parameter $h=0.71$ is assumed through out.\footnote{For
other values of $h$, our weak lensing masses scale as $M \propto h^{-1}$ and the X-ray
luminosities as $\Lx \propto h^{-2}$.}  In addition, the weak lensing data presented in this
analysis assumed a flat $\Lambda$CDM cosmology with $\Omega_m=0.27$.   
The recovered mass function has the standard hubble parameter
degeneracy.


\section{Data Sets}
\label{sec:data}

In this work we use the public maxBCG cluster catalog presented in \citet{koesteretal07a},
which is an optically selected volume limited catalog of close to $14,000$ clusters 
over the redshift range $z\in[0.1,0.3]$.  These clusters were found
in $~7500\ \deg^2$ of imaging data from the Sloan Digital Sky Survey \citep[SDSS,][]{yorketal00}
using the maxBCG cluster finding algorithm \citep{koesteretal07}.  This algorithm identifies
clusters as overdensities of red sequence galaxies. All clusters are assigned a redshift
based on the SDSS photometric data only, and these redshifts are known to be accurate 
to within a dispersion $\Delta z \approx 0.01$.  Every cluster is also assigned a richness
measure $N_{200}$, which is the number of red sequence galaxies above a luminosity cut
of $0.4L_*$ and within a
specified scaled aperture, centered on the Brightest Cluster Galaxy (BCG) of
each cluster.  Only clusters with $N_{200} \geq 10$ are included in the final catalog.
Interested readers are referred to \citet{koesteretal07a} and \citet{koesteretal07} for further 
details.  In the interest
of economy of notation, from now on we denote the maxBCG richness measure
simply as $N$.

The relationship between cluster richness and various well known mass tracers
has been studied in large, homogeneous samples, such as 2MASS \citep{daietal07}
and SDSS \citep{beckeretal07, 
johnstonetal07,rykoffetal08a,mandelbaumetal08}.  Of particular
interest to us are the weak lensing measurements of the mean mass as a function of richness, 
and the X-ray measurements of the mean and scatter of the
X-ray luminosity as a function of richness.   The former analysis has been carried out
by \citet{johnstonetal07} based on the weak lensing data presented in
\citet{sheldonetal07}, and independently by \citet{mandelbaumetal08b}.
In short, \citet{sheldonetal07} stacked maxBCG clusters 
within narrow richness bins, and measured the average weak lensing
shear profile of the clusters.
These shear profiles were turned into surface mass density contrast profiles
using the redshift distribution of background sources estimated with the methods
of \citet{limaetal08} and the neural net photometric redshift estimators described
in \citet{oyaizuetal08}. Then,
\citet{johnstonetal07} fit the resulting profiles using a halo model scheme 
to obtain tight constraints on the mean mass of maxBCG clusters for 
each of the richness bins under consideration.  
The \citet{mandelbaumetal08} analysis is very similar in spirit to the one described above.
The main differences are the way the source redshift distribution is estimated, and 
the details of the model fitting use to recover the masses.  The differences in
the results between these two analysis are discussed in appendix \ref{sec:m-n_priors},
where we use them to set priors on the mass--richness relation.

The measurement of the mean X-ray luminosity of maxBCG clusters has
been carried out by \cite{rykoffetal08a} following an approach similar to that pioneered
in \citet{daietal07}.  The necessary X-ray data is readily available from the ROSAT All-Sky
Survey \citep[RASS,][]{vogesetal99}.  In short, \cite{rykoffetal08a} stacked
the RASS photon maps \citep{vogesetal01} centered on
maxBCG clusters in narrow richness bins.  The background subtracted 
stacked photon counts within a $750\ h^{-1}\ \kpc$ aperture were used to 
estimate the mean X-ray luminosity $\Lx$ in the $0.1-2.4\ \keV$ rest frame
of the clusters.  In addition, \cite{rykoffetal08a} measured the scatter in
X-ray luminosity at fixed richness by individually measuring $\Lx$ for
all maxBCG clusters with $N\geq 30$.  It is worth noting that due to the 
shallowness of RASS,  many of the maxBCG clusters are not X-ray 
luminous enough to be detected individually.  However, non-detection
and upper limits for $\Lx$ for individual systems 
were properly taken into consideration using the
Bayesian approach detailed in \citet{kelly07}, and the recovered mean
X-ray luminosity from this Baysian analysis was fully consistent
with the stacked means.

In addition to the data sets above, we use the constraints on the $\Lx-M$ relation from 
\citet{vikhlininetal08}.  These constraints are based on the 400d cluster X-ray survey, a flux
limited cluster survey based on ROSAT pointed observations with an effective
sky coverage of 397 $\deg^2$ \citep{bureninetal07}.  Briefly, \citet{vikhlininetal08} measured 
both the total soft band X-ray luminosity and the cluster
mass for each cluster in the sample.  X-ray luminosities are estimated from
ROSAT data, and measure the luminosity in the rest-fram $0.5-2.0\ \keV$ band,
extrapolated to infinity assuming standard $\beta$ profiles.
Cluster masses are estimated based
on the values of $Y_X$ derived from followup Chandra observations, though they note
that the results they obtain using different mass tracers such as X-ray temperature and
total gas mass are very similar.  The $M-Y_X$ relation is
itself calibrated based on hydro-static mass estimates.   Importantly, \citet{vikhlininetal08}
explicitly correct for the Malmquist bias expected for a flux limited cluster sample, so the
$\Lx-M$ relation they derive can be interpreted as the relation one would obtain using a
mass limited cluster sample.

For this work, we have repeated the analysis in \citet{rykoffetal08a}
with a slightly different definition for $\Lx$.  In particular, we measure
the X-ray luminosity in the rest-frame $0.5-2.0\ \keV$ band within a $1\ h^{-1} \Mpc$
aperture.  The change in band is tailored to match the energy band
used by \citet{vikhlininetal08} , which we used to place priors on the 
$\Lx-M$ relation.  It is worth noting that \citet{vikhlininetal08} do not
use a $1\ h^{-1}\Mpc$ aperture, as we do.  We have, however, carefully
calibrated the scaling between our $\Lx$ definition and that of \citet{vikhlininetal08}
so as to be able to use their results in our analysis.end
A detailed description of  our measurements can be found in appendix \ref{app:lx-n_priors}.


\section{Relating Cluster Mass, X-ray Luminosity, and Richness}
\label{sec:rough}

The problem we are confronted with is the following: we have four pieces of observational
data, namely
\begin{itemize}
\item The abundance of galaxy clusters as a function of richness.
\item The mean relation between cluster richness and mass.
\item The mean and variance of the relation between cluster richness and
X-ray luminosity.
\item The mean and variance of the relation between cluster X-ray luminosity and
mass.
\end{itemize}
From this data, we wish to determine the scatter in mass
at fixed richness for the cluster sample under consideration.

The basic idea behind our analysis is as follows. Consider the
probability $P(M,L_X|N)$, which we take to be Gaussian in $\ln M$ and $\ln \Lx$.  
This probability distribution is completely specified
by the mean and variance of both $M$ and $\Lx$ at fixed richness, and by the
correlation coefficient between $M$ and $\Lx$.   Of these, there are only two
quantities that are not already observationally constrained: $\sigma_{M|N}$, 
the scatter in mass
at fixed richness, and $r_{M,L|N}$, the correlation coefficient between mass and $\Lx$ at fixed richness.

Suppose now that we guessed values for these two quantities, so that
the probability distribution $P(M,\Lx|N)$ is fully specified.  Given the 
abundance function $n(N)$,
we can use $P(M,\Lx|N)$ to randomly assign a mass and an X-ray
luminosity to every cluster in the sample.  We can then select a mass
limited sub-sample, and measure the corresponding
$\Lx-M$ relation, comparing it to the $\Lx-M$ measurement from \citet[][]{vikhlininetal08}.
Since the $\Lx-M$ relation
we predict depends on our assumptions about $P(M,\Lx|N)$,
there should only be a small region in parameters space where
our predictions are consistent with independent observational constraints on the
$\Lx-M$ relation.


\begin{figure}[t]
\epsscale{1.2}
\plotone{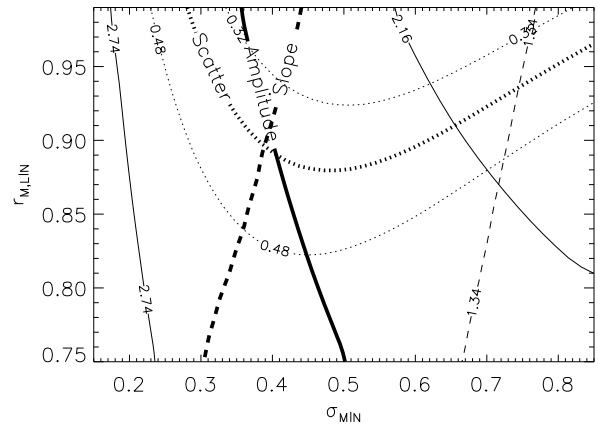}
\caption{Contours of constant $\Lx-M$ parameters.  For each assumed
value of the scatter $\sigma_{M|N}$ and correlation coefficient
parameter $r_{M,L|N}$, we predict the amplitude, slope, and scatter
of the $\Lx-M$ relation of a mass selected sample of clusters with
$M\geq 3\times 10^{14}\ \msun$.
Contours of constant amplitude, slope, and scatter are shown with
the solid, dashed, and dotted lines respectively.  The thicker lines
correspond to the central values of the $\Lx-M$ priors discussed in appendix
\ref{app:lx-m_priors} and summarized in Table \ref{tab:priors}, 
while the the other two contours enclose
the $95\%$ confidence region for each of the parameters.  The second slope contour
falls to outside the region of parameter space shown in the figure.
The intersection of the three separate regions correspond to acceptable 
values for the two unknown parameters $\sigma_{M|N}$ and $r_{M,L|N}$.
}
\label{fig:contours}
\end{figure} 


Figure \ref{fig:contours} illustrates this idea.  To create the figure, we have
set every observed parameter of the distribution $P(M,\Lx|N)$ to the
central value of the priors described in appendix \ref{app:priors} and summarized
in table \ref{tab:priors}.   We then
defined a grid in the two dimensional space spanned by $\sigma_{M|N}$
and $r_{M,L|N}$, and carried through the argument described above.
The resulting predictions for the amplitude, slope, and scatter of the
$\Lx-M$ relation as a function of $\sigma_{M|N}$ and $r_{M,L|N}$
are shown in the figure.
We plot contours of constant amplitude, slope, and scatter of the $\Lx-M$
relation as solid, dashed, and dotted lines respectively.  The thicker curves
correspond to the central values of the priors,
while thinner curves demark
the corresponding $95\%$ confidence limits. 
As we can see, all three contours intersect in a finite
region of parameter space, indicating good agreement between our weak lensing
and X-ray data, and the independent determination of the $\Lx-M$ relation.   
Based on Figure \ref{fig:contours}, we expect a detailed analysis should constrain
our parameters to
$\sigma_{M|N}\approx 0.40$, and $r_{M,L|N}\approx 0.9$.  
The rest of this paper is simply a way of formalizing the argument described 
above in order to place errors on both $\sigma_{M|N}$ and $r_{M,L|N}$.


\section{Formalism}
\label{sec:formalism}

We wish to formalize the above argument in order to place quantitative
constraints on the scatter in mass at fixed richness.  Details of how we go 
about doing so are presented below.  Readers interested only in our results
can move directly to section \ref{sec:results}.

\subsection{Likelihood Model}

As we mentioned above,
the key point in our analysis is our ability to compute 
the amplitude and slope of the mean relation $\avg{\ln \Lx|M}$, 
and the scatter about this mean, as a function of our two parameters
of interest: the scatter in mass at fixed richness and the correlation 
coefficient between $M$ and $\Lx$ at fixed $N$.  
Let us define $\bx=\{ A_{L|M},\alpha_{L|M}, \sigma_{L|M} \}$, and let
$\bp=\{ \sigma_{M|N}, r_{M,L|N} \}$ denote our parameters of interest.
Our predictions for the $\Lx-M$ relation as a function of our parameters
of interest can be summarized simply as $\bx(\bp)$.
Now, adopting a Bayesian framework, a set of priors on $\bx$ is simply a probability
distribution $P_\bx(\bx)$.   Since $\bx$ is a function of $\bp$, the priors immediately define
a probability distribution over $\bp$ given by
\begin{equation}
P(\bp) = P_\bx(\bx(\bp))\det (\partial \bx/\partial \bp).
\end{equation}
Since we know how to compute both $P_\bx(\bx)$ and $\bx(\bp)$,
we can find any confidence regions for our parameters of interest.

The problem we are confronted with, however, is slightly more complicated, in that
the functions $\bx$ depend not only on $\bp$, but also on additional nuisance parameters $\bq$.
Indeed, our predictions for
the observable parameters of the $\Lx-M$ relation depend on both the abundance
function of clusters and $P(M,\Lx|N)$.  The abundance function can
be accurately described by a Schechter function (we explicitly checked a Schechter function is
statistically acceptable),
\begin{equation}
n(N) \propto N^{-\tau}\exp( - N/N_* ).
\label{eq:abundance}
\end{equation}
Given a Schechter fit, 
our prediction for the $\Lx-M$ relation will also depend on the value
of the parameters $\tau$ and $N_*$.  Likewise, the distribution $P(M,\Lx|N)$
also depends on the
amplitude and slope of the means $\avg{M|N}$ and $\avg{\Lx|N}$, as well as
the scatter in $\Lx$ at fixed $N$.  All in all, we have six additional
nuisance parameters
$\bq=\{ N_*, \tau, B_{M|N},\alpha_{M|N}, A_{L|N}, \alpha_{L|N}, \sigma_{L|N} \}$.
Let $\br=\{\bp,\bq\}$ denote the full set of parameters.  The priors
from the $\Lx-M$ relation define a probability distribution over $\br$ given by
\begin{equation}
P(\br) = P_\bx(\bx(\br))\det (\partial \bx/\partial \br).
\end{equation}
Since we have a total of 8 parameters, and only three observables from the $\Lx-M$ relation, 
it is obvious that the above likelihood function will result in large degeneracies because
the parameters are under-constrained.
If one has priors $P_0(\bq)$ in the nuisance parameters, however,
the probability distribution $P(\bp)$ in the parameters of interest is given by
\begin{equation}
P(\bp) = \int d\bq\ P_0(\bq) P_\bx(\bx(\bp,\bq))\det(\partial\bx/\partial \br).
\label{eq:prob}
\end{equation}
This equation allows us to compute $P(\bp)$, and therefore place constraints
on our parameters of interest.
In practice, we will ignore the determinant term in the probability distribution defined
in equation \ref{eq:prob}.  This is because the function $\bx(\br)$ is estimated using
a Monte Carlo approach, implying that accurate numerical estimates of the 
Jacobian $\partial \bx/\partial \br$ would be too computationally intensive to
be performed.  Fortunately, the determinant typically introduces only 
slight modulations of the likelihood, so we do not expect our results to
be adversely affected by this.  


\subsection{Implementation}
\label{sec:implementation}

We estimate the probability distribution $P(\bp)$ using a Monte Carlo approach.
Ignoring an overall normalization constant and setting 
$\det (\partial \bx/\partial \br)=constant$, we have
\begin{equation}
\hat P(\bp) = \frac{1}{N_{draws}} \sum_{i=1}^{N_{draws}} P_\bx(\bx(\bp,\bq_i))
\label{eq:probestimator}
\end{equation}
where $\bq_i$ for $i=1$ through $N_{draws}$ are random draws of the
nuisance parameters $\bq_i$, drawn from the prior distribution $P_0(\bq_i)$.
We set $N_{draws}=3000$ as our default value (see below for further discussion).

The prior distributions for our nuisance parameters are characterized
by a statistical and a systematic error.  The former is modeled as 
Gaussian and the latter using a top-hat distribution.  Thus, given a prior
of the form
\begin{equation}
q=\bar q \pm \sigma_q^{stat} \pm \sigma_q^{sys},
\end{equation}
a random draw is obtained by setting
\begin{equation}
\bq_i = \bar \bq + \Delta \bq_i^{stat} + \Delta \bq_i^{sys}
\end{equation}
where $\Delta \bq_i^{stat}$ is drawn from a Gaussian of zero mean with a covariance matrix
defined by the statistical errors, and $\Delta \bq_i^{sys}$ is drawn from a top hat distribution
that is non-zero only for $|\Delta q^{sys} | \leq \sigma_q^{sys}$.

The probability distribution $P_\bx(\bx(\bp,\bq))$ used in equation 
\ref{eq:probestimator} is the product of the likelihoods $P_{x}(x(\bp,\bq))$
for each of the $\Lx-M$ parameters $x \in \bx=\{ A_{L|M}, \alpha_{L|M},\sigma_{L|M} \}$.
The probability for each $\Lx-M$ parameter is given by 
the convolution of the top-hat and Gaussian distributions defined by the statistical and systematic
errors of $x$, so that
\begin{equation}
P_{x}(x(\bp,\bq)) = \frac{1}{4\sigma_x^{sys}} [ \erf(x_+) - \erf(x_-) ]
\end{equation}
where
\begin{equation}
x_\pm = \frac{ \pm \sigma_x^{sys} - ( x(\bp,\bq) - \bar x ) }{\sqrt{2}\sigma_x^{stat}}.
\end{equation}
Note that the above equations are appropriate only when the various $\Lx-M$
parameters are uncorrelated, so it is important to place the priors at the pivot
point of the $\Lx-M$ relation ($M_{pivot}=3.9\times10^{14}\ \msun$).  
This explains why Table \ref{tab:priors} quotes 
a prior on $A_{L|M}+1.361\alpha_{L|M}+1.5(\sigma_{L|M}^2-0.40^2)$ rather than
on $A_{L|M}$ alone.

We also need to specify how the function $\bx(\bp,\bq)$ is evaluated.  We do this
using a Monte Carlo approach.  Given $\bp$ and $\bq$, we generate $N_{cl}=10^5$ 
mock clusters in the richness range
$N\in[10,200]$.
We then randomly draw mass and X-ray luminosity values for each of these clusters based on 
the distribution $P(M,\Lx|N)$, and select a mass limited subsample of clusters using a mass
cut $M\geq M_{min}$ with
$M_{min}=3\times 10^{14} \msun$ (the reason for this particular value is explained below).
Using a least squares fitting routine, we find the best fit line between
$\ln \Lx$ and $\ln M$.  This defines both $A_{L|M}(\bp,\bq)$ and 
$\alpha_{L|M}(\bp,\bq)$.  The scatter $\sigma_{L|M}(\bp,\bq)$ is defined
as the root mean square fluctuation about the best fit line.  

Using equation \ref{eq:probestimator} and the function $\bx(\bp,\bq)$ defined
above, we evaluate the probability distribution $P(\bp)$ along a grid of points in 
$\sigma_{M|N}\in[0.2,0.85]$ and $r\in[0.75,1.0]$ with $25$ grid points per axis.
A full run of our code then requires we perform $25^2$ Monte Carlo integrals with
$N_{draws}=3000$ points in each integration.  Each draw also requires us to
evaluate the function $\bx(\bp,\bq)$, which in turn requires generating a 
mock catalog with $N_{cl}=10^5$ clusters, so the procedure as a whole
is computationally expensive.
To increase computational efficiency, for each Monte Carlo evaluation of $P(\bp)$
we generate a single cluster catalog that is used to estimate the likelihood 
at every grid point.  This correlates the values of $\hat P$ along our grid,
but does not otherwise adversely affect our results.

Our Monte Carlo approach requires that
both the number of clusters in the random catalogs $N_{cl}$ and the
number of times the likelihood function is evaluated $N_{draws}$ is 
sufficiently large to achieve convergence.   Our default values for $N_{cl}$ and $N_{draws}$
were selected to ensure the recovered likelihood is accurate to
within a dispersion of $\sim 1-2\%$ inside high likelihood regions.
The error in the recovered likelihood increases with decreasing likelihood,
but even in the tails of the distributions our estimates are accurate to about $10\%$.
This was explicitly tested by running a coarse grid with our default values for $N_{draws}$ 
and $N_{cl}$, and by repeating the analysis with both of these parameters increased by a factor 
of two.\footnote{It is worth noting that in order to create Figure \ref{fig:contours},
one needs to generate cluster catalogs with $N_{cl}\gtrsim 10^7$ clusters in order
for the contours to appear smooth by eye.  However, $N_{cl}=10^5$ is a sufficient number of 
clusters for our analysis,
since we only require that the noise in the likelihood be much smaller
than the width of the priors. Since the latter are quite wide, even relatively noisy estimates
of the $\Lx-M$ relation are sufficient for constraining the marginalized distribution.}

Finally, we emphasize that it is necessary to explicitly check whether our results are 
sensitive to the $N\geq 10$ cut applied to the maxBCG clusters sample.  In particular,
when selecting a mass limited subsample of clusters, we need to ensure that the mass
limit $M_{min}$ be sufficiently large that the number of clusters with $N\leq 10$ and
$M\geq M_{min}$ is insignificant.  We have explicitly checked that for our adopted
low mass cut $M_{min}\geq 3\times 10^{14}\ \msun$ our results are robust to the 
richness cut $N\geq 10$ by repeating the analysis in a coarse grid using
an $N\geq 8$ richness cut instead.  We find that the likelihood
estimates in both cases are in agreement to within the expected accuracy
of our Monte Carlo approach.


\subsection{Priors}
\label{sec:priors}

The priors used in our analysis 
are summarized in Table \ref{tab:priors}.  We follow the notation
\begin{equation}
q = \bar q \pm \sigma_q^{stat}\ (stat) \pm \sigma_q^{sys}\ (sys)
\end{equation}
where $\bar q$ is the central value, $\sigma_q^{stat}$ is the 1$\sigma$ statistical error
on the parameter $q$ marginalized over all other parameters, and $\sigma_q^{sys}$
is the systematic error.  In all cases,
we model statistical errors as Gaussian, and we include known covariances between
different parameters.  Systematic errors are assumed to follow top-hat distributions, and
the final prior distribution is given by the convolution of these two functions.

\begin{deluxetable}{|c|c|}
\tablecaption{Scaling Relation and Cluster Abundance Priors}
\startdata
\hline
\hline
Parameter & Prior \\
\hline
$\ln N_*$ & \hspace{0.0 in} $3.66 \pm 0.10\ (stat) \pm 0.01\ (sys)$\hspace{0.0 in} \\
\hline
$\tau$ & \hspace{0.0 in} $2.61 \pm 0.06\ (stat)\pm 0.05\ (sys) $ \hspace{0.0 in} \\
\hline
$B_{M|N}$ & \hspace{0.0 in} $0.95 \pm 0.07\ (stat) \pm 0.10\ (sys)$  \hspace{0.0 in} \\
\hline
$\alpha_{M|N}$ & \hspace{0.0 in} $1.06 \pm 0.08\ (stat) \pm 0.08\ (sys)$   \hspace{0.0 in} \\
\hline
$B_{L|N}$ & \hspace{0.0 in} $1.91\pm 0.04\ (stat) \pm 0.09\ (sys)$\hspace{0.0 in} \\
\hline
$\alpha_{L|N}$ & \hspace{0.0 in} $1.63\pm 0.06\ (stat) \pm 0.05\ (sys)$  \hspace{0.0 in} \\
\hline
$\sigma_{L|N}$ & \hspace{0.0 in} $0.83\pm 0.03\ (stat) \pm 0.10\ (sys)$ \hspace{0.0 in} \\
\hline
\hline
\hspace{0.05 in} $A_{L|M}+1.361\alpha_{L|M}  +1.5(\sigma_{L|M}^2-0.40^2)  $ \hspace{0.05 in}  
	& \hspace{0.0 in} $2.45 \pm 0.08\ (stat) \pm 0.23\ (sys) $ \hspace{0.0 in} \\
\hline
$\alpha_{L|M}$ & \hspace{0.0 in} $1.61 \pm 0.14\ (stat)$   \hspace{0.0 in} \\
\hline
$\sigma_{L|M}$ & \hspace{0.0 in} $0.40 \pm 0.04\ (stat)$  \hspace{0.0 in} 
\enddata
\tablenotetext{}{Priors on the abundance function parameters ($N_*$ and $\tau$), as well as those
from the $M-N$ and $\Lx-N$ relations are not taken directly from any single work in the literature,
but are discussed in detail in Appendix \ref{app:priors}.  Priors on the $\Lx-M$ relation are taken
from \citet{vikhlininetal08}.  Overall, we believe these priors are fair, that is, they are neither
overly optimistic nor overly pessimistic.}
\label{tab:priors}
\end{deluxetable}

We believe that the priors contained in table \ref{tab:priors} are fair, that is, they are neither
overly aggressive nor overly conservative.  A detailed discussion of our priors
can be found in appendix \ref{app:priors}.


\section{Results}
\label{sec:results}

\begin{figure}[t]
\epsscale{1.2}
\plotone{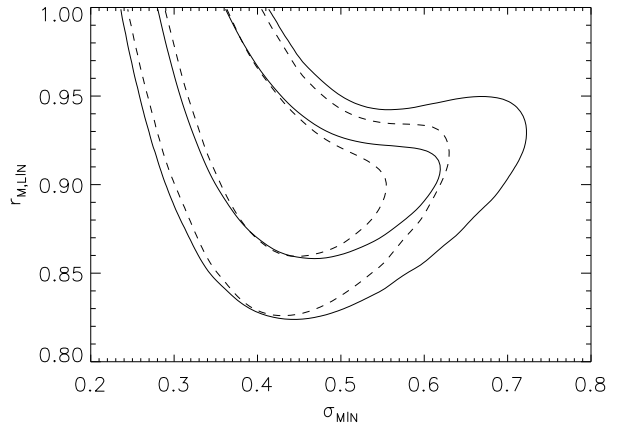}
\caption{$68\%$ and $95\%$ confidence contours for $\sigma_{M|N}$ and $r_{L,M|N}$. 
Solid lines show the results of our analysis.   We find that
X-ray luminosity and mass are correlated at fixed richness.  The breadth of
the degeneracy region shown above is almost exclusively due to uncertainties
in the $\Lx-M$ relation parameters.  Dashed contours demonstrate how our
results would improve if the $\Lx-M$ amplitude and slope were known to within
an accuracy of $\Delta A_{L|M} = \Delta \alpha_{L|M}=0.05$.}
\label{fig:lkhd}
\end{figure}

Figure \ref{fig:lkhd} shows the $68\%$ and $95\%$ probability contours for the parameters
$\sigma_{M|N}$ and $r_{M,L|N}$.   The likelihood peak occurs at
$\sigma_{M|N}=0.46$ and $r_{M,L|N}=0.90$.   The marginalized means
are $\avg{\sigma_{M|N}}=0.45$ and $\avg{r_{M,L|N}}=0.91$.

We wish to determine whether the breadth of the likelihood region in Figure \ref{fig:lkhd}
is limited by uncertainties in the scaling relations
of maxBCG clusters, or by uncertainties in the $\Lx-M$ relation.  To do so, we repeat
our analysis with two new sets of priors: for the first, we use a tight $0.05$ statistical
prior on all nuisance parameters, but let the $\Lx-M$ parameters float.  The second set of priors 
uses a tight $0.05$ prior on each of the  
$\Lx-M$ parameters, but floats all other nuisance parameters with the original priors.
We find that using tight priors on our nuisance parameters has negligible impact on 
the likelihood regions recovered from our analysis.  
On the other hand, the confidence regions obtained with the tight $\Lx-M$ priors, 
shown in Figure \ref{fig:lkhd} as dashed curves,
are tighter than those derived from our original analysis.
Thus, the dominant source of error in our analysis is the uncertainty in the values of the
$\Lx-M$ parameters.  This can be easily understood based on Figure \ref{fig:contours}.
We can see from the figure that the uncertainty in $r_{M,L|N}$ is largely due to
the prior on the scatter in $\Lx$ at fixed $M$, which is already tight and thus does not change
between our fiducial prior and our tight priors.   On the other hand, we can see that both
the amplitude and slope priors cut-off regions with high scatter.  Tightening
these priors excludes a larger section of parameter space, and results in the tighter contours observed
in Figure \ref{fig:lkhd}.


\begin{figure}[t]
\epsscale{1.2}
\plotone{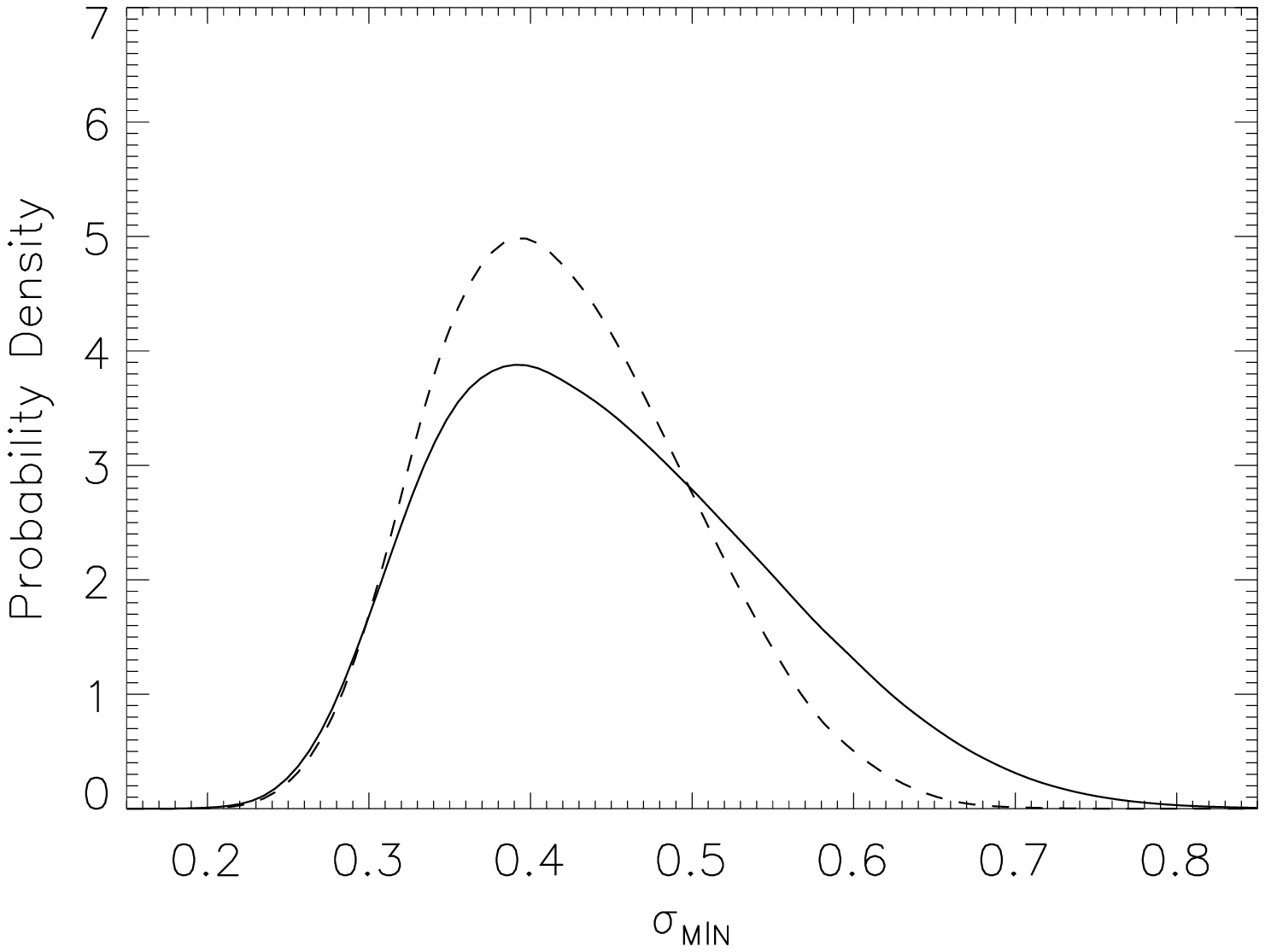}
\plotone{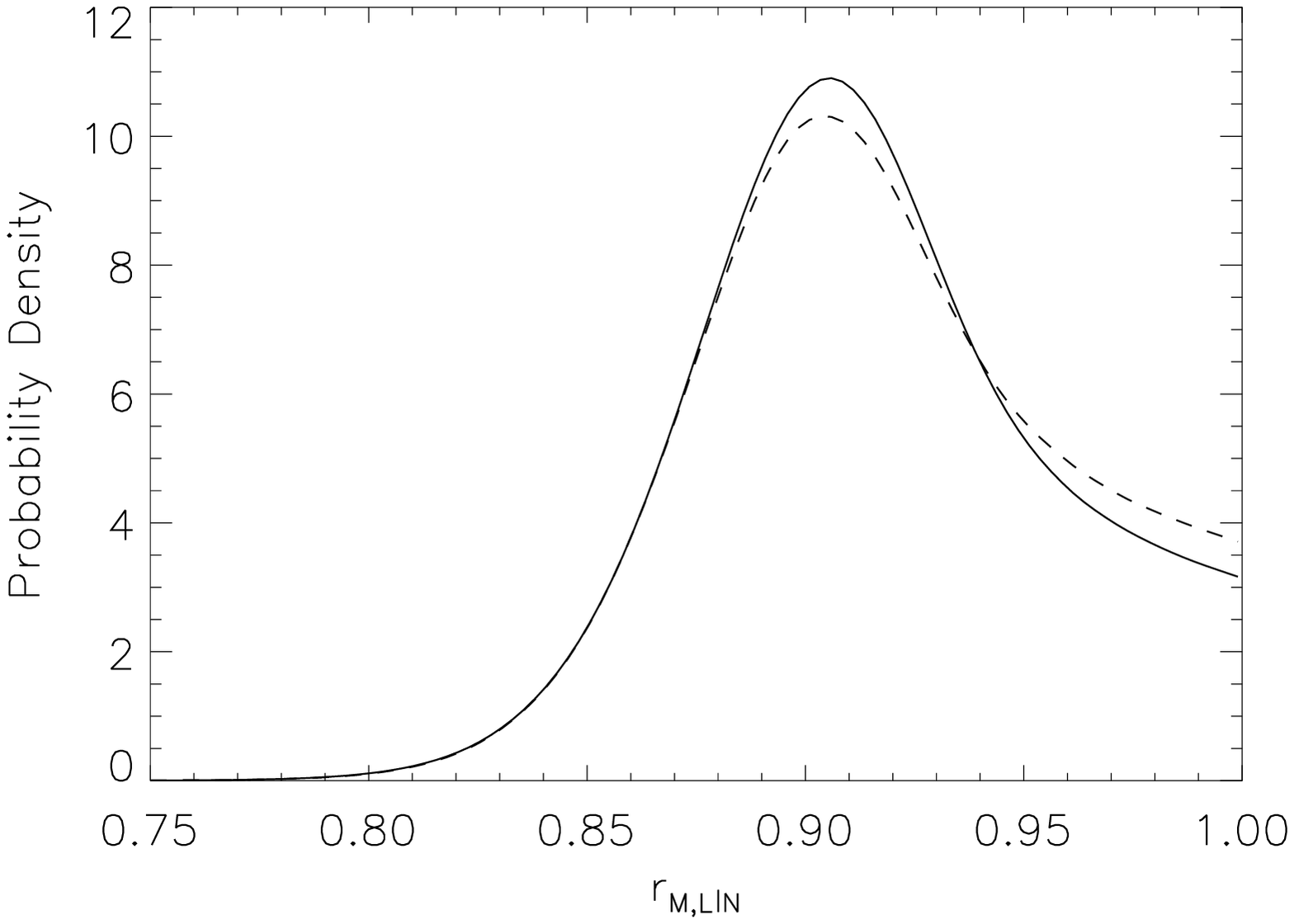}
\caption{Likelihood distributions for $\sigma_{M|N}$ and $r_{M,L|N}$.  The distributions
are marginalized over all other parameters.  Solid lines are the results of our analysis,
while dashed lines are the results obtained assuming tight priors on the $\Lx-M$ parameters
Note the latter set of curves are presented 
only to give a sense of how our result would improve with
better understanding of the $\Lx-M$ relation.}
\label{fig:marg}
\end{figure}

Figure \ref{fig:marg} shows the marginalized probability distributions for 
$\sigma_{M|N}$ and $r_{M,L|N}$.
The solid curves correspond to our original analysis, while the dashed curves illustrate
the results one expects assuming our hypothetical tight priors for the $\Lx-M$ relation parameters.
We find that the logarithmic scatter in mass at fixed richness
and the correlation coefficient between $\ln M$ and $\ln \Lx$ are
\begin{eqnarray}
\sigma_{M|N} & = & 0.45^{+0.20}_{-0.18}\ (95\%\ \mbox{CL}) \\
r_{L,M|N} & \geq & 0.85\ (95\%\ \mbox{CL}).
\end{eqnarray}
Assuming our hypothetical tight $\Lx-M$ priors, the constraints
become $\sigma_{M|N} = 0.42^{+0.07}_{-0.09}$ and 
$r_{L,M|N} \geq 0.85\ (95\%\ \mbox{CL})$.
We emphasize that these latter constraints are only meant as a guide to 
the accuracy one could achieve with this method if the $\Lx-M$
relation were known to about $5\%$ accuracy.

It is evident from our results that cluster richness is not as effective a
mass tracer as X-ray derived masses. Indeed, even total (i.e. not core-core excluded) 
X-ray luminosity is a more faithful mass tracer than
the adopted richness measure of the maxBCG catalog, as demonstrated both 
by the smaller scatter and the very large correlation
coefficient. Note that the latter indicates that, at fixed richness, over-luminous
clusters are almost guaranteed to also be more massive than average.  This
is an important result which forms the basis for a concurrant paper in which we improve our richness
estimates by demanding tighter correlations in the $\Lx-$richness relation~\citep{rozoetal08b}.


\section{Comparison to Other Work}
\label{sec:other_work}

There are not many previous results against which our measurements of scatter in mass 
at fixed richness may be compared.
One possible reference point is the upper limit based on
the error bar in the weak lensing mass estimates of \citet{johnstonetal07}.  More specifically,
assuming that the error in $\avg{M|N}$ is entirely due to the intrinsic scatter
in $M$ at fixed $N$, it follows that the error in the mass is simply 
$\Delta M/\avg{M|N} \approx \Delta \ln M = \sigma_{M|N}/\sqrt{n(N)}$ where $\Delta M$ is the observed error
and $n(N)$ is the number of clusters with richness $N$.  
For the richest bin, which provides the tightest constraint, 
\citet{johnstonetal07} find 
$\avg{M}=(8.1\pm 1.3)\times 10^{14}\ \msun$.  The bin contains $n=47$
clusters, so an upper limit to the scatter in mass at fixed richness is
$\sigma_{M|N} \leq \sqrt{n}(\Delta M/\avg{M})=1.10$.  Figure \ref{fig:marg}
shows that our results easily satisfy this upper limit on the scatter.

The only other measurement of the scatter in mass at fixed richness for maxBCG clusters
is that found in \citet{beckeretal07}.  These scatter estimates
are obtained as follows: first, \citet{beckeretal07} select all maxBCG clusters whose
central galaxy has a spectroscopic redshift.  They then bin the clusters in richness,
and compute the velocity relative
to the BCG of every galaxy member with spectroscopic data.  The recovered velocity
distribution of galaxies is found to be non-Gaussian.  Assuming that the velocity distribution
of galaxies of halos of fixed mass is exactly Gaussian, and that the observed non-Gaussianity
is entirely due to mass-mixing within a richness bin, \citet{beckeretal07} estimate the scatter in mass 
at fixed richness based on the observed non-Gaussianity of the velocity distribution.

\begin{figure}[t]
\epsscale{1.2}
\plotone{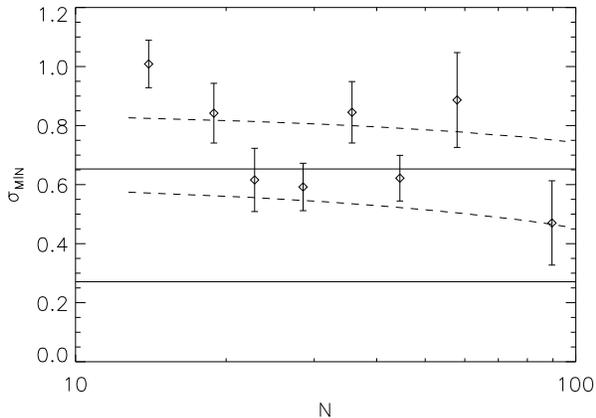}
\caption{Comparison of the scatter in mass at fixed richness estimated
in this work (solid band) and that of \citet{beckeretal07} (diamonds with
error bars).  The dashed band
shows how the scatter we measured is expected to be affected by miscentering,
which allows us better compare our results to those of \citet{beckeretal07}.
We find that, once 
miscentering is properly taken into account, the two results appear to be
in reasonable agreement.}
\label{fig:compare}
\end{figure}

An updated version of the results from \citet{beckeretal07} can be seen in 
Figure \ref{fig:compare}.  The only difference between this plot and the corresponding
figure in \citet{beckeretal07} is that here we have made used of the additional spectroscopic
data from the SDSS Data Release 6 \citep{dr6}, which results in tighter error bars.  
Also shown in the figure as a horizontal
band is the $95\%$ confidence region from our analysis.  As we can see, our scatter 
estimate appears to be systematically lower than that of \citet{beckeretal07}, a discrepancy
first noted in \citet[][more on the relation between our work and theirs below]{rykoffetal08b}.


Such a bias is not entirely unexpected, as we now know that a significant fraction
of cluster have their BCGs miss-identified, a problem that was not
yet known -- and was therefore unaccounted for -- at the time the 
\citet{beckeretal07} results came out.
To get a better understanding of how our results and those of \citet{beckeretal07}
compare, we can use our results along with the miscentering probability model 
from \citet{johnstonetal07} to predict the scatter that \citet{beckeretal07} observed
given this miscentering systematic.  We proceed as follows.  First, we
use our best fit model for the abundance distribution 
to generate a mock catalog with $2\times 10^5$ clusters with $N\geq 10$.
Each of these clusters is assigned a mass by drawing from the $P(M|N)$
distribution defined by the values of $\sigma_{M|N}$ corresponding
to the two $95\%$ confidence limits on $\sigma_{M|N}$.
These assigned masses
are then turned into velocity dispersions using the scaling relation from \citet{evrardetal08}.

At this point, we have a cluster catalog where each cluster has a richness and a velocity
dispersion.  If a cluster is miscentered, we expect that in most cases the new center
will be a cluster galaxy.  Assuming this is the case, and that 
BCGs are at rest at the center of a cluster, 
the velocity dispersion of cluster galaxies relative to random satellites will be a
factor of $\sqrt{2}$ high than relative to the BCG.
Using the miscentering model described in \citet{johnstonetal07} for $p(N)$, the probability 
that a cluster of richness $N$ be correctly centered, 
we randomly label clusters as properly centered or miscentered, and boost their 
"observed" velocity dispersion for those clusters labeled as miscentered by the expected amount.
The clusters are assigned a new mass based on their
``observed'' velocity dispersions, and the corresponding scatter in the 
$M-N$ relation is estimated.  We repeat this procedure $10^3$ times
in order to compute the mean systematic correction due to miscentering.

Our predictions for the scatter values observed by \citet{beckeretal07} are
shown in Figure \ref{fig:compare} with dashed lines, and correspond to the $95\%$
confidence interval from our analysis.  
We see that miscentering introduces a richness dependent correction that boosts 
the scatter in the recovered velocity dispersion  and places it in significantly better
agreement with the data from \citet{beckeretal07}.  

The agreement with the \citet{beckeretal07} data is an interesting
result.  Perhaps the single most difficult systematic effect that had to be addressed in the
\citet{beckeretal07} analysis is the validity of the assumption that non-Gaussianities in the
velocity distribution of stacked clusters are entirely due to mass-mixing is a valid.  
The reasonable agreement between our results and those of \citet{beckeretal07}
suggests that their assumption is indeed justified, though a robust conclusion will
have to wait until a more detailed analysis is performed, especially given the possibility
of velocity bias of the galaxy population (i.e. if satellite galaxies have a velocity dispersion different
from that of the dark matter).

The analysis in this work is also very closely related to that of \citet{rykoffetal08b}.  \citet{rykoffetal08b}
sought to constrain the $\Lx-M$ relation of clusters by fitting the scaling of $\avg{\Lx|N}$ with
$\avg{M|N}$.  However, as recognized in \citet{rykoffetal08b},
in order to fully interpret their result in terms of the traditional definition of the $\Lx-M$ relation, i.e. the mean
X-ray luminosity at fixed mass, one needs to know both the scatter in mass at fixed richness, and the corresponding
correlation coefficient with $\Lx$.  
Given that these two quantities are unknown, but that the $\Lx-M$ relation is already constrained from X-ray surveys, 
it seems reasonable to suggest that a better use of the lensing and X-ray data of maxBCG clusters is to 
use our knowledge of the $\Lx-M$ relation to constrain the scatter in mass at fixed richness and the corresponding
correlation coefficient, as was done in this work.

Our work differs from the ideas presented in \citet{rykoffetal08b} in another significant way.   While our analysis
employs only $P(\Lx,M|N)$ and $n(N)$, \citet{rykoffetal08b} used the halo mass function $dn/dM$ and the
probability distribution $P(\Lx,N|M)$ to interpret their measurements.  This has the important drawback that
in doing so, one needs to assume a cosmological model in order to compute the halo mass function, rendering
their interpretation cosmology dependent.  By focusing on the quantities that are directly observable, i.e.
$n(N)$ and $P(\Lx,M|N)$, we are able to avoid this difficulty.   The price we pay for this is that rather than constraining
the scatter in richness at fixed mass, which is the more directly relevant quantity from a cosmological perspective,
we constrain instead the scatter in mass at fixed richness.  While this makes implementing
such a constraint a little more cumbersome in a cosmological analysis, the fact that the constraint itself is cosmology
independent is obviously of paramount importance.  


\section{Cosmological Consequences}
\label{sec:mf}

As mentioned in the introduction, to obtain an unbiased estimate of the halo mass function
based on the observed cluster richness function requires that we understand the scatter between
cluster richness and halo mass.
Given our lognormal assumption, and the fact that the mean
mass--richness relation is already known from weak lensing, our measurement of the scatter in this
scaling relation fully determines the probability distribution $P(M|N)$.  Thus, 
we are now in a position to determine the halo mass function of the local universe 
with the maxBCG cluster catalog.

Let us define then
$n_i=n(M_i)$ as the number of halos within a logarithmic mass bin of width
$\Delta \ln M$ centered about $M_i$,
\begin{equation}
n_i = \left . \frac{dn}{d\ln M} \right \vert_{M_i} \Delta \ln M.
\end{equation}
Given our cluster catalog and $P(M|N)$, we construct an estimator
$\hat n_i$ for $n_i$ by randomly drawing a mass from $P(M|N)$ 
for each halo in the cluster catalog, and then counting the number of halos
within the logarithmic mass bin centered about $M_i$.  
Note that since the mass of each cluster is a random variable, our
mass function estimator $\hat n_i$ is itself a random
variable.  The mean and correlation matrix of $\hat n_i$ can easily be obtained by making multiple
realizations of $\hat n_i$, and averaging the resulting mass functions.

In practice, we also need to marginalize our results over uncertainties
in $P(M|N)$ and over uncertainties in the richness function $n(N)$.
To do so, we randomly draw the parameters
$\bx=\{B_{M|N},\alpha_{M|N},\sigma_{M|N}\}$, and then resample of the
cluster richness function to obtain
a new estimate of $n_i$.  The whole procedure is iterated $10^5$ times,
and the mean and covariance matrix of the number counts
in each of our logarithmic mass bins is computed.\footnote{We again checked
explicitly that the mass cut $M_{min}=3\times 10^{14}\ \msun$ is large enough
for our results to be insensitive to the maxBCG richness cut $N\geq 10$.}  


\begin{figure}[t]
\epsscale{1.2}
\plotone{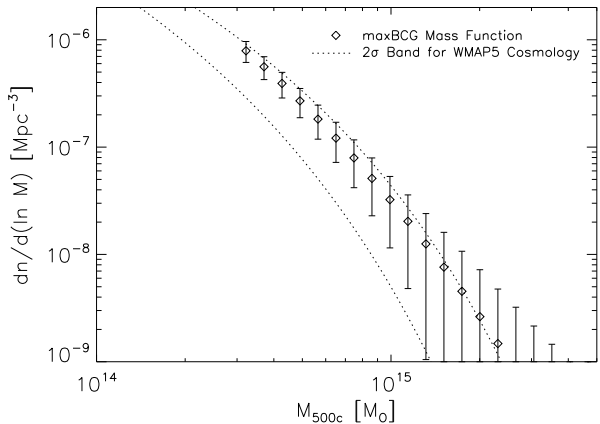}
\caption{The maxBCG mass function.  Cluster counts were converted
to densities assuming $\Omega_m=0.27$ and $h=0.71$, the same cosmology
assumed in the lensing measurements \citep{johnstonetal07}.
The error bars shown are due to
the scatter in the mass--richness relation, and are strongly correlated.
For comparison, we have
also plotted the \citet{tinkeretal08} mass function corresponding to the
WMAP5 $95\%$ confidence region for $\sigma_8$,
$0.724 \leq \sigma_8 \leq 0.868$.  All other parameters are held fixed to 
the central values reported in \citet{wmap08}.  Our data are consistent with
the WMAP5 results, though they might suggest a slightly higher power spectrum
normalization.}
\label{fig:mf}
\end{figure} 


Figure \ref{fig:mf} shows the mass function recovered through our
analysis.  To turn our number counts into a density, we assumed
a WMAP5 cosmology \citep{wmap08}, with $\Omega_m=0.27$
and $h=0.72$, and a photometric redshift error $\Delta z=0.01$ (used for
computing the effective volume of the sample).
The diamonds correspond to our estimated means, and
the error bars are the square root of the diagonal elements of
the correlation matrix.  We emphasize that the error bars are heavily
correlated.  The mean and covariance matrix of the recovered halo mass function
can be found in Appendix \ref{app:mfdata}.

Also shown in Figure \ref{fig:mf} with dotted lines are the halo mass functions at
$z=0.23$ predicted by WMAP5 assuming the \citet{tinkeretal08} mass function.
For both curves, we set all cosmological parameters to the central values reported
in \citet{wmap08}, except for $\sigma_8$, which is set to $\sigma_8=0.868$
for the upper curve and $\sigma_8=0.724$ for the lower curve.  These two
values define the $95\%$ confidence interval for $\sigma_8$ in \citet{wmap08}.
As we can see, the mass function recovered from our analysis is fully consistent
with the WMAP5 cosmology, though it seems to push for values of $\sigma_8$ on the
high end of their allowed region.  A detailed cosmological analysis
of our data will be presented in a subsequent paper (Rozo et al, in preparation).


\section{Summary and Conclusions}
\label{sec:conclusions}

We have shown that by combining the information in the maxBCG
richness function, the mean richness-mass relation, the
mean and scatter of the $\Lx-$richness relation, and the mean and scatter
of the $\Lx-M$ relation, we can constrain both the scatter in mass
at fixed richness for maxBCG clusters, as well as the correlation coefficient
between mass and $\Lx$ at fixed richness.  We find
\begin{eqnarray}
\sigma_{M|N} & = & 0.45^{+0.20}_{-0.18}\ (95\%\ \mbox{CL}) \\
r_{L,M|N} & \geq & 0.85\ (95\%\ \mbox{CL}).
\end{eqnarray}
These constraints are dominated by uncertainties in the $\Lx-M$ relation, and can be significantly
tightened if our understanding of the $\Lx-M$ relation improves.  We also found our
results are consistent with those presented in \citet{beckeretal07} once miscentering
of maxBCG clusters is taken into account.

Our lower limit on the correlation between $M$ and $\Lx$ at fixed richness constitutes the first observational
constraint on a correlation coefficient involving two different halo mass tracers.  Note that the large
correlation between $\Lx$ and $M$ implies that $\Lx$ - even without core exclusion - is a significantly
better mass tracer than the maxBCG richness estimator (i.e. at fixed richness, over-luminous cluster
are nearly always more massive).  This is an important result, which we use in
a concurrent paper to help us define new richness estimators that are better correlated with
cluster mass~\citep{rozoetal08b}.

Using our results, and assuming $\Omega_m=0.27$ and $h=0.71$,
we have estimated the halo mass function at $z=0.23$, corresponding to the median redshift of the
cluster sample.  We find that our recovered mass function is in good agreement with 
the mass function predicted by \citet{tinkeretal08} for the WMAP5 cosmology \citep{wmap08}. A
detailed cosmological analysis will be presented in a forthcoming
paper (Rozo et al, in preparation).

Our work sheds new light on the interrelationship of bulk properties of massive halos.
We have used weak lensing, X-ray luminosities,
and optical richness estimates to constrain the scatter in the richness-mass relation, which can lead
to improved cosmological constraints.  In principle, one could also turn this question around, and, assuming
cosmology, we could constrain the scatter in the richness-mass relation, which would then allow us
to place constraints on the amplitude, slope, and scatter of the $\Lx-M$ relation.   Such an analysis would
be interesting in that, by doing so, one could compare the predicted amplitude of the $\Lx-M$ relation to
that derived from hydrostatic mass estimates, thereby directly probing the amount of non-thermal pressure
support in galaxy clusters. Note that even though this question can also be directly addressed by comparing
weak lensing and X-ray mass estimates of individual clusters, the analysis suggested here would benefit
from having small uncertainties, whereas projection effects result in rather noisy weak lensing
mass estimates for individual systems.  

\acknowledgements The authors would like to thank Alexey Vikhlinin for providing them with 
the full covariance of the $\Lx-M$ parameters from their analysis of the 400d cluster sample.  
ER thanks David Weinberg and Chris Kochanek for useful discussions and their careful reading
of the manuscript.  ESR would like to thank the TABASGO foundation.
TM and JH gratefully acknowledge support from NSF grant AST 0807304 and DoE 
Grant DE-FG02-95ER40899.  AE gratefully acknowledges support from NSF grant 
AST-0708150.   RHW was supported in part by the U.S. Department of Energy under contract
number DE-AC02-76SF00515 and by a Terman Fellowship at Stanford
University.
This project was made possible by workshop support from the Michigan Center for Theoretical Physics.


\newcommand\AAA[3]{{A\& A} {\bf #1}, #2 (#3)}
\newcommand\PhysRep[3]{{Physics Reports} {\bf #1}, #2 (#3)}
\newcommand\ApJ[3]{ {ApJ} {\bf #1}, #2 (#3) }
\newcommand\PhysRevD[3]{ {Phys. Rev. D} {\bf #1}, #2 (#3) }
\newcommand\PhysRevLet[3]{ {Physics Review Letters} {\bf #1}, #2 (#3) }
\newcommand\MNRAS[3]{{MNRAS} {\bf #1}, #2 (#3)}
\newcommand\PhysLet[3]{{Physics Letters} {\bf B#1}, #2 (#3)}
\newcommand\AJ[3]{ {AJ} {\bf #1}, #2 (#3) }
\newcommand\aph{astro-ph/}
\newcommand\AREVAA[3]{{Ann. Rev. A.\& A.} {\bf #1}, #2 (#3)}

\appendix


\section{Priors}
\label{app:priors}

\subsection{Abundance Priors}

Our estimates of the $\Lx-M$ parameters depend on the abundance function
of maxBCG clusters, which is observationally determined, but not known to 
infinite precision.  Here, we fit the observed abundance function using a Schechter
function, such that the mean number of clusters $\mu$ of richness $N$ is 
\begin{equation}
\mu(N) = n_0 (N/40)^{-\tau} \exp( - N/N_* ).
\end{equation}
The amplitude $n_0$ is chosen such that the total number of clusters exactly
equals the observed number of clusters.  We set this normalization condition
because we are interested only in the shape of the richness function, and not 
in its amplitude.  

The fits are done by maximizing the likelihood of the observed distribution,
binned in bins of width $\Delta N=1$.  We assume that the probability
distribution of observed $n$ clusters in a bin of richness $N$ is Poisson,
with
\begin{equation}
P(n) = \frac{ \mu(N)^n\exp(-\mu(N)) }{n!}.
\end{equation}
For numerical purposes, we cut the distribution at $N_{max}=300$,
which is sufficiently large to not affect our fits.  We emphasize that we use the
above likelihood only to define estimators for $N_*$ and $\tau$, since, as discussed
below, both goodness of fit and errors in the parameter estimation are obtained through
Monte Carlo simulation.

The richness distribution is fit over the range $N\geq 10$ by maximizing the
log-likelihood function using an amoeba routine.  To estimate our
errors, we follow a Monte Carlo approach and resample the observed richness 
function $10^4$ times.  We find that the parameters $N_*$ and $\tau$ are significantly
correlated, with the probability distribution being Gaussian in 
$\tau$ and $\ln N_*$.   The best fit parameters are
\begin{eqnarray}
\avg{\ln N_*} & = & 3.66\pm 0.10 \\
\avg{\tau} & = & 2.61 \pm 0.06
\end{eqnarray}
with a correlation coefficient 
\begin{equation}
r_{N_*,\tau} = 0.94.
\end{equation}

To assess goodness of fit, we generate $10^4$ mock catalogs
with as many clusters as the real data from the probability distribution specified by
$\avg{\ln N_*}$ and $\avg{\tau}$.  We compute the likelihood for each of these mock catalogs,
and compare the corresponding likelihood distribution to that observed in the real data.
We find that our fit is statistically acceptable.

The most significant systematic error affecting our measurements of the shape
of the richness function are completeness and purity variations in the cluster catalog.  
\citet{rozoetal07b} have shown that the maxBCG catalog
is over $90\%$ pure and complete for $N\geq 10$.  Here, we take a conservative approach,
and consider the change in
the best fit parameters assuming the observed counts are rescaled by a completeness/purity
correction factor $\lambda$ given by
\begin{equation}
\lambda=\min \{ 0.9+0.1\ln(N/10.0)/\ln(10.0)\}.
\end{equation}
This corresponds to a $10\%$ decrease in the observed counts at $N=10$ while
holding the counts at $N=100$ constant.  
Upon refitting the data after this correction we find systematic offsets
\begin{eqnarray}
(\Delta \ln N_*)_{sys} & = & 0.01 \\
(\Delta \tau)_{sys} & = & 0.05
\end{eqnarray}
which we adopt as our systematic error.  Note the systematic offsets are allowed
to be both positive and negative, since the correction multiplier $\lambda$ above
could easily be larger than unity rather than smaller than unity.


\subsection{$M-N$ Priors}
\label{sec:m-n_priors}

Our priors on the $M-N$ relation are based on the results presented in \citet{johnstonetal07},
\citet{mandelbaumetal08}, and \citet{mandelbaumetal08b}.  To assign our priors, we first 
compare the results of these two works as a means of assessing systematic uncertainties
in the mass parameters.  We then focus exclusively on the \citet{johnstonetal07} results
to place our final priors on the $M-N$ relation.  The latter choice reflects the fact that
\citet{johnstonetal07} report weak lensing mass estimates for several mass definitions, among
them $M_{500c}$, the relevant quantity in the $\Lx-M$ relation of \citet{vikhlininetal08}.

Let us then begin by discussing the \citet{johnstonetal07} results first.  
While \citet{johnstonetal07} quote a power-law
fit for the mean mass at fixed $\avg{M|N}$, this fit is based a non-public version of the
maxBCG catalog that extends 
to a richness of $N=3$ (the catalog for clusters with $N<10$ is not public).   Since the maxBCG
catalog is only known to be highly complete and pure in the range $N\geq 10$, we have refit
the \citet{johnstonetal07} masses restricting ourselves to the range $N\geq 9$.  This slightly
lower cut is necessary due to the richness binning in \citet{johnstonetal07}.    We find that the
mass $M_{180b}$ within a 180 overdensity threshold relative to mean matter density is 
\begin{equation}
\frac{\avg{M_{180b}|N}}{10^{14}\ h^{-1}\msun} = \exp(0.25\pm0.07)(N/20)^{1.18\pm0.09}
\end{equation}
with a correlation coefficient $r=-0.43$ between the amplitude and slope parameters.  

\citet{mandelbaumetal08b} preformed a similar but independent weak lensing analysis of the
maxBCG clusters, though using $M_{200b}$ as their mass variable.  They find
\begin{equation}
\frac{\avg{M_{200b}|N}}{10^{14}\ h^{-1} \msun} = \exp(0.45\pm 0.08)(N/20)^{1.15\pm 0.14}
\end{equation}
To compare against the \citet{johnstonetal07} values, we use the \citet{hukravtsov03} mass 
conversion formulae to find an approximate power law relation between $M_{200b}$ and $M_{180b}$
over the range $5\times 10^{14}\ \msun \leq M_{200b} \leq 10^{15}\ \msun$.  We find
$M_{180b} = 1.022 M_{200b}$, which is only a $2\%$ correction.  Applying this correction,
we find that the corresponding $M-N$ parameters from \citet{mandelbaumetal08b} are
\begin{equation}
\frac{\avg{M_{180b}|N}}{10^{14}\ h^{-1}\msun} = \exp(0.47\pm0.07)(N/20)^{1.15\pm0.14}
\end{equation}

We find that the slopes of the \citet{johnstonetal07} and \cite{mandelbaumetal08b} results are 
nearly identical, but that the masses of \citet{mandelbaumetal08b} are
systematically higher by $\approx 25\%$.  This difference can be traced back to how
the lensing critical surface density for each of the two works is estimated. 

In general, lensing masses are proportional to the quantity
$1/\avg{ \Sigma_{crit}^{-1} }$, where $\Sigma_{crit}$ is the lensing critical
surface density, and the average is to be computed over the source redshift
distribution. Given multi-band photometric data $\bm{m}$ for each
galaxy, one way to compute $\avg{\Sigma_{crit}^{-1}}$ is to use a photometric
redshift estimator $z_{photo}(\bm{m})$, and then assume that the true
source redshift distribution is identical to the photometric redshift distribution.
\citet{mandelbaumetal08} have shown that such a simple
approach typically results in biased lensing mass estimates, but they also
demonstrate that it is possible to achieve unbiased results using the probability
distribution $P(z|\bm{m})$.

The weak lensing analysis in \citet{sheldonetal07}, on which the results from \citet{johnstonetal07}
are based, falls somewhere in between these two approaches.  While \citet{sheldonetal07} 
does in fact make use of photometric redshifts, they do not simply assume that the source
redshift distribution is identical to the photometric redshift distribution.  Rather, they construct
a probability distribution $P(z|z_{photo})$, and use this probability to 
estimate $\avg{\Sigma_{crit}^{-1}}$.
As it turns out, evaluating $\avg{\Sigma_{crit}^{-1}}$ in this way
leads to results that are nearly identical to those obtained by simply setting $z=z_{photo}$.
Thus, even though the approach used in
\citet{sheldonetal07} is more sophisticated than the simple case considered
in \citet{mandelbaumetal08}, we expect the \citet{sheldonetal07} results to be biased but correctable
as prescribed in \citet{mandelbaumetal08}.
This correction amounts to a boost of the lensing masses by a factor of $1.18\pm 0.02\ (stat) \pm 0.02\ (sys)$.
The statistical error bar in the correction is added in quadrature to the statistical error
bar from our fit, which results in
\begin{equation}
\frac{\avg{M_{180b}|N}}{10^{14}\ h^{-1}\msun} = \exp[0.42 \pm 0.07\ (stat) \pm 0.02\ (sys)]\times (N/20)^{1.18\pm 0.09}
\label{eq:johnston_mass}
\end{equation}
These new values for the \citet{johnstonetal07} data are in considerably better agreement with those
of \citet{mandelbaumetal08b}.  There remains, however, a systematic $5\%$ difference between the
two amplitudes, as well as a small difference $\Delta \alpha_{M|N}=0.028$ between the two slopes.

A possible culprit for this systematic $5\%$ offset is the difference in how miscentering is accounted for
in the data models.
The word miscentering refers to the fact that when finding clusters, one will inevitably find clusters that are
improperly centered, either due to failures of the cluster finding algorithm, or simply because there is
no obvious center of the cluster based on its optical image.  Such offsets between the true and assigned
centers are problematic because if a cluster is miscentered, the corresponding lensing
signal is weakened, resulting in systematically low mass estimates.  

To determine whether the remaining offset between \citet{mandelbaumetal08b} and \citet{johnstonetal07}
is consistent with differences in the miscentering model, we refit our data assuming no errors
on the miscentering corrections.  We find
\begin{equation}
\avg{M_{180b}|N} = \exp(0.42\pm 0.04)(N/20)^{1.17\pm 0.07}.
\end{equation}
with a correlation coefficient $r=-0.15$. Note that these errors are smaller than the errors
quoted before, as they should be, given that this new fit does not marginalize over a wide range
of miscentering models.  By subtracting the two sets of errors in quadrature, we find
that the miscentering priors adopted in \citet{johnstonetal07} correspond to an error
$0.043$ in the amplitude and $0.05$ in the slope.  Thus, the \citet{mandelbaumetal08b}
mass measurements are well within the centering error included in the analysis 
of \citet{johnstonetal07}.

Nevertheless, it is unclear whether miscentering can in fact account for the difference between
the \citet{johnstonetal07} and \citet{mandelbaumetal08} results.  More specifically, 
\citet{mandelbaumetal08b} also performed their analysis including the \citet{johnstonetal07}
model for miscentering, and find after applying the centering correction their best fit
$M_{180b}-N$ relation becomes
\begin{equation}
\avg{M_{180b}|N} = \exp(0.53\pm 0.07)(N/20)^{1.08\pm 0.14}.
\end{equation}
Comparing this to equation \ref{eq:johnston_mass}, we find including a miscentering correction in
the \citet{mandelbaumetal08b} analysis increases the tension between the two results.  Moreover,
it suggests that the difference between the two results is due to some other form of systematic
difference between the two analysis pipelines.  
In light of this, we opt for introducing a systematic correction to the \citet{johnstonetal07}
results of  $+0.06$ and $-0.05$ for the amplitude and slope respectively.  We also introduce 
systematic errors of the same magnitude as this systematic correction, so that our final result is
\begin{equation}
\avg{M_{180b}|N} = [\exp(0.48\pm 0.07\ (stat) \pm 0.06\ (sys))](N/20)^{1.13\pm 0.09\ (stat) \pm 0.05\ (sys)}.
\end{equation}
Note the central values of the original \citet{johnstonetal07} analysis (corrected for photometric redshift bias)
as well as the \citet{mandelbaumetal08b} results both with and without miscentering corrections are all encompassed
by our systematic error.  

Now, in this work we are interested more in the $M_{500c}-N$ (henceforth simply $M-N$) relation than in the 
$M_{200c}-N$ relation, since it is the former mass which is accessible to X-ray studies.  To constrain 
the $M-N$ relation we use the quoted $M_{500c}$ mass measurements from \citet{johnstonetal07}, re-scaling their
$M_{200c}$ errors to $M_{500c}$ by assuming the relative errors are constant.  A fit to the data results in
\begin{equation}
\frac{\avg{M|N}}{10^{14} \msun} = \exp[ 0.68 \pm 0.07 ] (N/40)^{1.11 \pm 0.08}
\end{equation}
with a correlation coefficient $r=0.45$.  

We now boost this expression by factor $1.18$ due to the photometric redshift bias correction, and add the systematic
corrections $+0.06$ and $-0.05$ to the amplitude and slope respectively as per our discussion of the $M_{180b}-N$ relation.
We also include a systematic error on the amplitudes and slopes of this same magnitude.  We obtain
\begin{eqnarray}
B_{M|N} & = & 0.91 \pm 0.07\ (stat)\ \pm 0.06\ (sys) \\
\alpha_{M|N} & = & 1.06 \pm 0.08\ (stat) \pm 0.05\ (sys).
\end{eqnarray}

The final systematics we consider here are the purity and completeness of the sample.   Now,
as long as the completeness is not correlated with mass, completeness should not in any
way bias the recovered parameters of the $M-N$ relation, though it obviously affects the
error bars due to lower statistics.  

The same cannot be said of purity.   If only a fraction $p$ of the clusters are actually
good matches to real halos in the universe, then a fraction $1-p$ of the clusters will have a lensing
signal that is significantly different from the mean signal.  As an extreme case, we can consider
what happens if a fraction $1-p$ of the clusters had no mass associated with them.  In that case,
the observed mean mass is simply $M_{obs}=M_{true}/p$ where $M_{true}$ is the true mean, 
so one should boost the observed masses
by a factor of $1/p$ to obtain an unbiased estimate. For $p=0.9$, this amounts to an 
increase in $B_{M|N}$ of magnitude  $\Delta B_{M|N}= 0.1$. 
Now, \citet{rozoetal06} showed that the purity of the maxBCG cluster sample
is expected to be above $90\%$ over the range or richnesses considered here,
and the increase in $B_{M|N}$ quoted above is undoubtedly an overestimate of the
necessary correction since even false cluster detections will have excess mass
associated with them.  In light of this, we have adopted a one-sided systematic error bar 
$\Delta B_{M|N}=0.08$ to take into account the impact of purity in the recovered $M-N$
relation.   The error bar is one sided since we expect impurities will tend to decrease
the observed mean mass.  We can, however, turn this prior into a normal double-sided
prior by including a systematic correction $\Delta B_{M|N}=0.04$ to the central
value, and setting the systematic error bar to the same magnitude as the
central value shift.
We can also get a rough estimate for the systematic error on the purity 
by assuming that the quoted systematic error in the amplitude should be made only
in the limit of high or low richness.  If that were the case, using the fact the slope is measured
over a decade of richness values, the corresponding slope
would be
\begin{equation}
\alpha = \frac{ 1.06\ln(10)+0.08}{\ln(10)} = 1.09
\end{equation}
which amounts to a systematic offset $\Delta\alpha=0.03$.  
These systematic error bars are added linearly to our previous systematic error.
Our final set of priors for the $M-N$ relation is
\begin{eqnarray}
B_{M|N} & = & 0.95 \pm 0.07\ (stat) \pm 0.10\ (sys) \\
\alpha_{M|N} & = & 1.06 \pm 0.08\ (stat) \pm 0.08\ (sys)
\end{eqnarray}
with a correlation coefficient $r=0.45$ between the two statistical errors.


\subsection{$\Lx-N$ Priors}
\label{app:lx-n_priors}

The priors in the $\Lx-N$ relation come from repeating the analysis described
in \citet{rykoffetal08a}, but with $\Lx$ defined as the X-ray luminosity in the
0.5-2.0 keV band, and corrected for aperture effects.  As in
\citet{rykoffetal08a}, we restrict this analysis to clusters with $N \geq
30$.  We begin by measuring the stacked mean $\Lx-N$ relation and scatter on a
fixed $1\,\hMpc$ scale
\begin{eqnarray}
\label{eqn:lxnmean}
B_{L|N} & = & 1.69\pm0.04\,(stat)\\
\alpha_{L|N} & = & 1.63 \pm 0.06\,(stat)\\
\sigma_{L|N} & = & 0.84 \pm 0.03\,(stat)
\end{eqnarray}
where we have measured $\Lx$ in units of
$10^{43}\ \mathrm{ergs}\times \mathrm{s}^{-1}$, with a pivot point of
$N = 40$.  We emphasize that the scatter determined above is the total
scatter in the observed $\Lx-N$ relation that cannot be attributed to Poisson
uncertainties in the ROSAT photon counts.  In particular, the quoted scatter is
affected by possible point source contamination, AGN activity, cool cores,
cluster mergers, etc.

There are multiple systematic errors that can affect the derived parameters for
the $\Lx-N$ relation.  These include photometric redshift errors,
evolution of the richness parameter $N$, uncorrelated point sources,
cluster mis-centering, and cluster AGN and cool cores.  In addition, we need to
account for the fraction of cluster flux lost due to our finite aperture and
the RASS PSF, in order to compare our results with the luminosity measurements
of \citet{vikhlininetal08}.  We shall now discuss each of these possible
systematic effects.

\citet{rykoffetal08a} find that the accuracy of the maxBCG photo-z estimates is
high enough such that any biases are insignificant relative to the statistical
uncertainty of the parameter determinations, and can thus be safely ignored.
However, \citet{rykoffetal08a} did find significant redshift evolution in the
$\Lx-N$ relation, well above the expected self-similar evolution.
Similar redshift evolution is found in \citet{beckeretal07}; the reason for the
systematic undercounting of cluster members at high redshift is explained in \citet{rozoetal08b}.  We have estimated the effect of this redshift evolution on
our derived scatter parameter via a simple Monte Carlo, and confirm that
although the apparent evolution is strong, it is insignificant relative to the
intrinsic scatter.  Therefore, we may also safely ignore this possible
systematic effect.

We now take a combined approach to the systematic effects due to cluster
mis-centering, a finite aperture, the RASS PSF and uncorrelated point sources.
The first three effects are strongly related, in that they all tend to scatter
cluster photons out of our initial fixed $1\,\hMpc$ aperture, and these may
affect the normalization, slope, and scatter in the $\Lx-N$ relation.
Uncorrelated point sources should not affect the mean relation because the
large number of stacked sources smooths out the foreground and background.
However, when uncorrelated point sources are aligned with individual clusters
they may increase the measured scatter by boosting the apparent $\Lx$.

We have estimated the effects of these systematics by running a Monte Carlo
with simulated RASS data on top of random backgrounds selected from the area of
the RASS photon map that overlaps with the maxBCG mask. 
We first resample the maxBCG richness function 100 times.  Each 
cluster is given a redshift drawn from the maxBCG redshift distribution, as well as a random 
postion on the sky selected from the area of the RASS survey that overlaps with the maxBCG mask.  
After we select the richest 1000 clusters in each realization, each cluster is given a luminosity based 
on the mean relation from Eqn.~\ref{eqn:lxnmean} and an input intrinsic scatter,  
$\sigma_{in} =\{0.0, 0.2, 0.4, 0.6, 0.8, 1.0\}$.  Each cluster luminosity is then converted
to a number of photon counts according to the RASS exposure at the given
point, and scattered by Poisson uncertainties.  Then, each cluster is given a
position offset according to the maxBCG miscentering distribution described in
\citet[][see \S~4.3]{johnstonetal07}.  The cluster profiles are assumed to
follow a $\beta$ model, $S(R) = S_0(1+R^2/R_C^2)^{-3\beta+1/2}$.  To ensure we
are on similar footing as \citet{vikhlininetal08}, we randomly assign each
cluster $\beta$ model parameters uniformly in the range $0.6<\beta<0.7$ and
$0.05<R_C<0.15\,\hMpc$.  Finally, the photons are scattered according to the
RASS PSF, following the method of \citet[][see \S~3.3.1]{rykoffetal08a}.  We
then calculate the stacked mean relation and scatter as described in
\citet{rykoffetal08a}.

\begin{figure}
\begin{center}
\rotatebox{270}{\scalebox{0.5}{\plotone{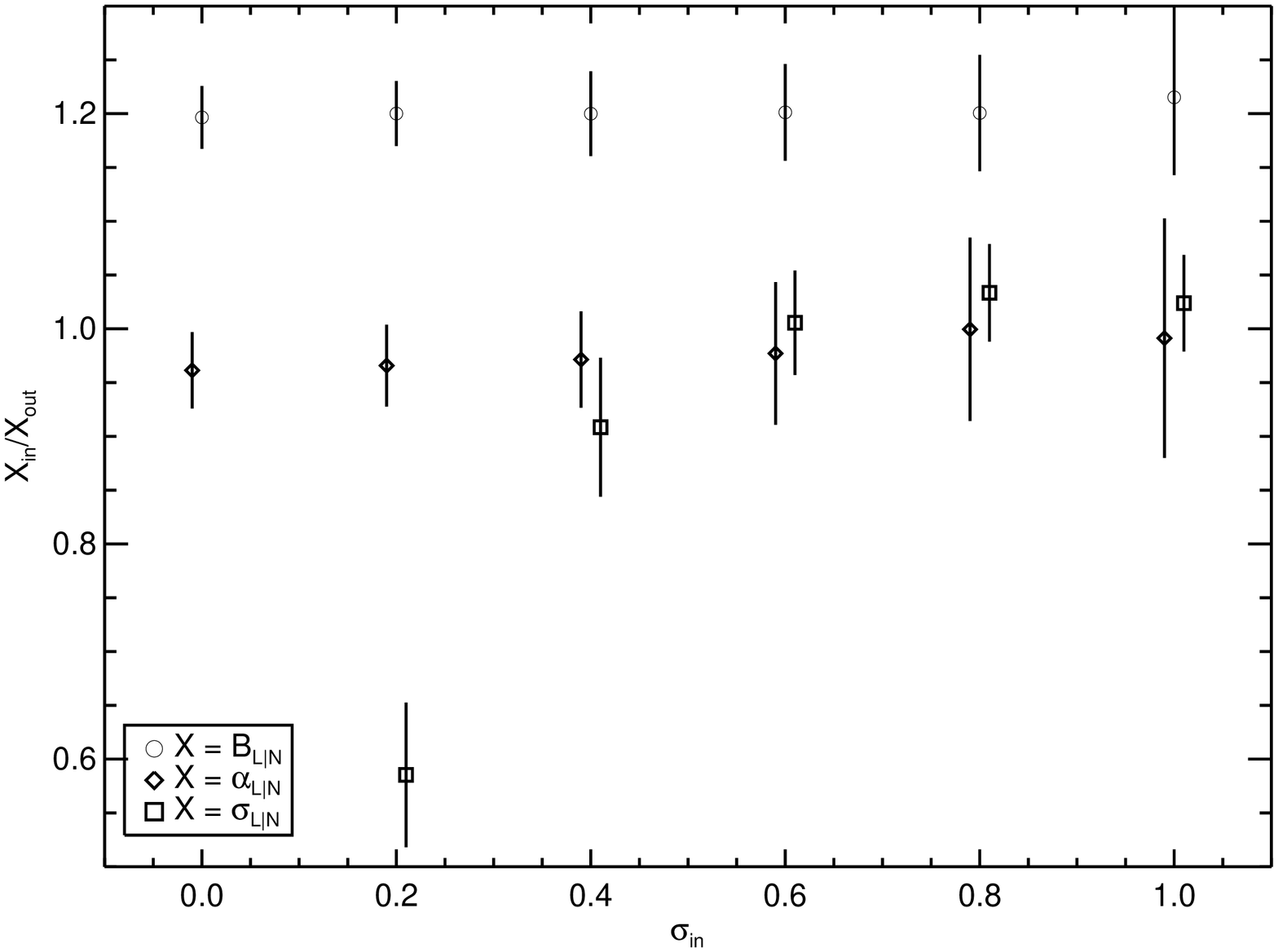}}}
\caption{\label{fig:lnsystematics}Results from systematic error Monte Carlo
  tests.  The x-axis shows the input intrinsic scatter, $\sigma_{in}$.  The
  y-axis shows the ratio of the given input parameter to output parameter for
  the normalization $B_{L|N}$ (circles), slope $\alpha_{L|N}$ (diamonds), and
  scatter $\sigma_{L|N}$ (squares).  We note that when $\sigma_{in}=0.0$ then
  $\sigma_{out}=0.31\pm 0.04$, which cannot be displayed on the plot.
}
\end{center}
\end{figure}

Figure~\ref{fig:lnsystematics} summarizes the results from our systematic
tests.  The x-axis shows the input intrinsic scatter, $\sigma_{in}$.  The
y-axis shows the ratio of the input parameter to output parameter for the
normalization $B_{L|N}$ (circles), slope $\alpha_{L|N}$ (diamonds), and scatter
$\sigma_{L|N}$ (squares).  We note that when $\sigma_{in}=0.0$ then
$\sigma_{out}=0.31\pm 0.04$, which cannot be displayed on the plot.  This is
consistent with our expectation that uncorrelated sources may boost the observed
scatter, especially with low intrinsic scatter.  Overall, we find that (a) the
slope $\alpha_{L|N}$ is not significantly biased; (b) at moderate to large
scatter ($\sigma_{in} \gtrsim 0.5$) the intrinsic scatter $\sigma_{L|N}$ is not
significantly biased; and (c) the output normalization $B_{L|N}$ must be
boosted by a factor of $1.20\pm0.05$ to account for the flux lost to
miscentering, the finite aperture, and RASS PSF effects.  Our priors become 
then
\begin{eqnarray}
B_{L|N} & = & 1.87 \pm 0.04\,(stat) \pm 0.05 \,(sys)\\
\alpha_{L|N} & = & 1.63 \pm 0.06\,(stat) \\
\sigma_{L|N} & = & 0.84 \pm 0.03\,(stat).
\end{eqnarray}

In addition to these corrections, we also need to take into account systematic uncertainties
due to purity and completeness in the sample.  Just as with the weak lensing mass estimates,
completeness should not affect the measured $\Lx-N$ relation, whereas purity will tend to
suppress the X-ray luminosity at fixed richness.  Following the same procedure as in 
appendix \ref{sec:m-n_priors}, we derive systematic errors $\Delta B_{L|N}=0.04$ and
$\Delta \alpha_{L|N}=0.05$, which we add linearly to our previous systematic error estimates.
Finally, we have repeated our scatter analysis using not just the 1000 richest clusters,
but also the 2000 richest clusters, in which case we find $\sigma_{L|N}=0.95$.  To 
take into account this variation in our analysis, we also introduce a systematic error 
$\Delta \sigma_{L|N}=0.10$.
Our final set of priors is
\begin{eqnarray}
B_{L|N} & = & 1.91 \pm 0.04\,(stat) \pm 0.09 \,(sys)\\
\alpha_{L|N} & = & 1.63 \pm 0.06\,(stat)\pm 0.05\,(sys)\\
\sigma_{L|N} & = & 0.84 \pm 0.03\,(stat)\pm 0.10\,(sys).
\end{eqnarray}

\subsection{$\Lx-M$ Priors}
\label{app:lx-m_priors}

As discussed in section \ref{sec:rough}, our analysis hinges on the fact
that we can use prior knowledge about the $\Lx-M$ relation 
to constrain the $M-N$ relation.   Here, we use the results of \citet{vikhlininetal08}
to put priors on the $\Lx-M$ relation, which may be summarized as\footnote{We have
included the appropriate evolution correction for a median redshift $z=0.23$, as
appropriate for the maxBCG sample.}
\begin{eqnarray}
A_{L|M} + 1.361\alpha_{L|M} +1.5(\sigma_{L|M}^2-0.40^2) & = & 2.59\pm 0.08 \\
\alpha_{L|M} & = & 1.61\pm 0.14 \\
\sigma_{L|M} & = & 0.40 \pm 0.04.
\end{eqnarray}
We report a prior on $A_{L|M}+1.361\alpha_{L|M}+1.5(\sigma_{L|M}^2-0.40^2)$ 
because at $M=10^{14}\ \msun$ the $\Lx-M$ parameters derived from the
\citet{bureninetal07} sample are correlated.  To decouple them, one
needs to shift to the statistical pivot point $M=3.9\times 10^{14}\ \msun$ and
introduce the scatter dependence quoted above (Vikhlinin, private communication).
These constraints are derived from Chandra observations of clusters in the
400d cluster catalog \citep{bureninetal07}, which allowed \citet{vikhlininetal08}
to measure $Y_X$ and thereby infer cluster mass using the $M-Y_X$ relation. 
This relation was itself calibrated on a cluster subsample for which masses were
derived using the standard hydrostatic equilibrium argument.
This last point is very important, since
simulations suggest that hydrostatic mass estimates of clusters may be biased low
by $\approx 10\%-30\%$ \citep[see e.g.][]{evrard90,rasiaetal06,nagaietal07a}.  
One way to calibrate such uncertainties is to
compare weak lensing mass estimates to hydrostatic mass estimates.   There are several
examples of this type of approach.  For instance, 
\citet{vikhlininetal08} have
performed such an analysis using the weak lensing mass estimates of \citet{hoekstra07}, and
find $M_{wl}= ( 1.09 \pm 0.11) M_X.$
A similar analysis has been carried out by \citet{mahdavietal08}, who used the weak lensing
mass estimates of \citet{hoekstra07} and their own analysis of Chandra public data to obtain
$M_{wl}=( 1.28 \pm 0.15) M_X$.  Finally, using XMM X-ray observations
and the weak lensing data of \citet{bardeauetal05}, \citet{bardeauetal07}, and \citet{dahle07},
\citet{zhangetal08} find $M_{wl} = ( 1.21 \pm 0.13 )M_X$.
\citet{zhangetal08} also note, however, that a histogram of $M_{wl}/M_{X}$ peaks
at a ratio of $1.00\pm 0.05$, and that clusters in the tails of the distribution tend to have tight
error bars, possibly biasing the error weighted ratio.
In light of this, we have opted for a ``middle of the road'' approach, and introduce a correction 
factor $1.15\pm 0.15$.  Our corresponding prior is
\begin{eqnarray}
A_{L|M} + 1.361\alpha_{L|M}  +1.5(\sigma_{L|M}^2-0.40^2)  & = & 2.45 \pm 0.08\ (stat) \pm 0.23\ (sys) \\
\alpha_{L|M} & = & 1.61\pm 0.14\ (stat) \\
\sigma_{L|M} & = & 0.40 \pm 0.04\ (stat).
\end{eqnarray}

Estimating systematic errors in $\alpha_{L|M}$ and $\sigma_{L|M}$ is difficult.  For instance,
comparisons with weak lensing masses are not an effective way of assessing systematics 
because weak lensing mass estimates are so noisy: trying to fit a power law relation between
$M_{wl}$ and $M_X$ results in very large errors for the slope of the relation.

One alternative is to consider multiple studies of the $\Lx-M$ relation in order to asses how
sensitive the recovered parameters are to the analysis pipeline.  Unfortunately, such an excercise
is far from trivial.  One difficulty is the fact that there is very little agreement on the meaning of
$\Lx$, with many works focusing on core-excised and/or core-corrected bolometric X-ray
lumunisoties \citep[e.g.][]{bardeauetal07b,zhangetal07,zhangetal08}.  Even among those works
that also explore the $\Lx-M$ relation when $\Lx$ is a soft X-ray band luminosity
\citep[e.g.][]{rb02,maughan07}, there are still important differences in the aperture used
to estimate $\Lx$.  In principle, we could attempt to convert between the various definitions
of $\Lx$ to try to compare the works against each other, but many of these $\Lx-M$ measurements
are affected by Malmquist bias, making comparisons to the \citet[][]{vikhlininetal08} results
difficult.

One work that does constrain the the soft X-ray band,
non-core excised, Malmquist bias corrected $\Lx-M$ relation is \citet{staneketal06}.
Unfortunately, the energy band they use is slightly different from that of of \citet{vikhlininetal08}, so
even here comparison is not trivial.  We expect, however, that at least the scatter and slopes 
of the $\Lx-M$ relation will not be strongly affected by the minor differences between the two 
$\Lx$ definitions.  Given our purposes, the interesting thing about the \citet{staneketal06}
results is that they use a very different methodology for constraining the $\Lx-M$ relation.
In particular, they assume knowledge of cosmological parameters, and then use the observed
cluster X-ray luminosity function to constrain $P(\Lx|M)$.  Assuming their ``compromise cosmology'',
which they argue gives the best results, they find
$\alpha_{X|M}=1.60\pm0.05$ and $\sigma_{L|M}=0.34\pm0.10$.  These values are in
excellent agreement with those of \citet{vikhlininetal08}, and suggest that placing additional
systematic errors in the $\Lx-M$ parameters is not really necessary at this point.


\section{Mass Function Data}
\label{app:mfdata}

Table \ref{tab:mfdata} presents the mean and covariance matrix of the mass function data derived from our
analysis.  These results represent the state of the art mass function measurements at low redshift
from optically derived cluster catalogs.  We emphasize we assumed
$\Omega_m=0.27$ and $h=0.71$, so appropriate rescaling must be applied if the results
are to be compared against significantly different cosmologies.  
Note that the covariance matrix data in table \ref{tab:mfdata} is 
normalized such that the diagonal entries are the fractional error $\sqrt{C_{i,i}}/\avg{n_i}$, while
the off diagonal entries are the correlation coefficients $r_{i,j}=C_{i,j}/\sqrt{C_{i,i}C_{j,j}}$.  We present
the data in this way since it is easier to understand when expressed this way.   The 
actual values for the covariance matrix are easily reconstructed from the data in the table.

\begin{deluxetable}{|c|c|lllllllllllllllll|}
\tablewidth{0pt}
\tablecaption{\label{tab:mfdata}maxBCG Mass Function Data}
\startdata
\hline
$M_{500c}$ & $\avg{dn/d\ln M}$ & 3.22 & 3.70 & 4.26 & 4.91 & 5.65 & 6.50 & 7.49 & 8.62 & 9.92 & 11.42 & 13.14 & 15.12 & 17.41 & 20.04 & 23.07 & 26.55 & 30.56 \\
\hline
\hline
3.22 & 7.90e-7 & 0.22 & 0.82 & 0.77 & 0.72 & 0.66 & 0.61 & 0.55 & 0.50 & 0.44 & 0.39 & 0.35 & 0.30 & 0.25 & 0.21 & 0.18 & 0.15 & 0.12 \\ 
3.70 & 5.61e-7 & 0.82 & 0.24 & 0.79 & 0.76 & 0.71 & 0.67 & 0.62 & 0.57 & 0.52 & 0.46 & 0.41 & 0.36 & 0.31 & 0.27 & 0.23 & 0.19 & 0.16 \\ 
4.26 & 3.92e-7 & 0.77 & 0.79 & 0.27 & 0.77 & 0.74 & 0.70 & 0.66 & 0.61 & 0.57 & 0.52 & 0.46 & 0.41 & 0.36 & 0.31 & 0.26 & 0.22 & 0.18 \\ 
4.91 & 2.70e-7 & 0.72 & 0.76 & 0.77 & 0.30 & 0.75 & 0.72 & 0.68 & 0.64 & 0.59 & 0.55 & 0.50 & 0.44 & 0.38 & 0.34 & 0.29 & 0.24 & 0.20 \\ 
5.65 & 1.82e-7 & 0.66 & 0.71 & 0.74 & 0.75 & 0.35 & 0.72 & 0.69 & 0.65 & 0.61 & 0.57 & 0.52 & 0.46 & 0.41 & 0.36 & 0.30 & 0.26 & 0.22 \\ 
6.50 & 1.21e-7 & 0.61 & 0.67 & 0.70 & 0.72 & 0.72 & 0.41 & 0.68 & 0.65 & 0.62 & 0.57 & 0.52 & 0.47 & 0.42 & 0.37 & 0.32 & 0.27 & 0.22 \\ 
7.49 & 7.93e-8 & 0.55 & 0.62 & 0.66 & 0.68 & 0.69 & 0.68 & 0.47 & 0.64 & 0.60 & 0.57 & 0.52 & 0.47 & 0.42 & 0.37 & 0.32 & 0.27 & 0.23 \\ 
8.62 & 5.11e-8 & 0.50 & 0.57 & 0.61 & 0.64 & 0.65 & 0.65 & 0.64 & 0.55 & 0.59 & 0.55 & 0.51 & 0.47 & 0.41 & 0.37 & 0.32 & 0.27 & 0.23 \\ 
9.92 & 3.24e-8 & 0.44 & 0.52 & 0.57 & 0.59 & 0.61 & 0.62 & 0.60 & 0.59 & 0.65 & 0.53 & 0.49 & 0.45 & 0.40 & 0.36 & 0.31 & 0.26 & 0.22 \\ 
11.42 & 2.03e-8 & 0.39 & 0.46 & 0.52 & 0.55 & 0.57 & 0.57 & 0.57 & 0.55 & 0.53 & 0.76 & 0.47 & 0.43 & 0.38 & 0.34 & 0.30 & 0.25 & 0.21 \\ 
13.14 & 1.25e-8 & 0.35 & 0.41 & 0.46 & 0.50 & 0.52 & 0.52 & 0.52 & 0.51 & 0.49 & 0.47 & 0.92 & 0.40 & 0.36 & 0.32 & 0.28 & 0.24 & 0.20 \\ 
15.12 & 7.61e-9 & 0.30 & 0.36 & 0.41 & 0.44 & 0.46 & 0.47 & 0.47 & 0.47 & 0.45 & 0.43 & 0.40 & 1.11 & 0.33 & 0.30 & 0.26 & 0.22 & 0.19 \\ 
17.41 & 4.53e-9 & 0.25 & 0.31 & 0.36 & 0.38 & 0.41 & 0.42 & 0.42 & 0.41 & 0.40 & 0.38 & 0.36 & 0.33 & 1.36 & 0.26 & 0.24 & 0.20 & 0.17 \\ 
20.04 & 2.63e-9 & 0.21 & 0.27 & 0.31 & 0.34 & 0.36 & 0.37 & 0.37 & 0.37 & 0.36 & 0.34 & 0.32 & 0.30 & 0.26 & 1.74 & 0.21 & 0.19 & 0.16 \\ 
23.07 & 1.48e-9 & 0.18 & 0.23 & 0.26 & 0.29 & 0.30 & 0.32 & 0.32 & 0.32 & 0.31 & 0.30 & 0.28 & 0.26 & 0.24 & 0.21 & 2.22 & 0.17 & 0.14 \\ 
26.55 & 8.29e-10 & 0.15 & 0.19 & 0.22 & 0.24 & 0.26 & 0.27 & 0.27 & 0.27 & 0.26 & 0.25 & 0.24 & 0.22 & 0.20 & 0.19 & 0.17 & 2.88 & 0.12 \\ 
30.56 & 4.41e-10 & 0.12 & 0.16 & 0.18 & 0.20 & 0.22 & 0.22 & 0.23 & 0.23 & 0.22 & 0.21 & 0.20 & 0.19 & 0.17 & 0.16 & 0.14 & 0.12 & 3.88 
\enddata
\tablenotetext{}{Mean and covariance matrix of the maxBCG mass function.   Masses are defined using an overdensity
of 500 relative to critical, and are measured in units of $10^{14}\ \msun$.  Space densities are measured in
units of $\Mpc^{-3}$.  Diagonal terms in the covariance matrix above are set to $\sqrt{C_{i,i}}/\avg{n_i}$, and thus represent
the fractional uncertainty in the halo space density.   Off diagonal terms contain the correlation coefficient $r_{i,j}=C_{i,j}/\sqrt{C_{i,i}C_{j,j}}$
between the various bins.  The median redshift of the sample is $z=0.23$.}
\end{deluxetable}

\end{document}